\documentclass[number,3p,onecolumn,preprint]{elsarticle}

\usepackage{algorithm}
\usepackage{algpseudocode}
\usepackage{amsfonts}
\usepackage{amsmath}
\usepackage{amssymb}
\usepackage{amsthm}
\usepackage{bbm}
\usepackage{booktabs}
\usepackage{dsfont}
\usepackage{mathrsfs}
\usepackage{multirow}
\usepackage{pgfplots}
\usepgfplotslibrary{fillbetween}
\usepackage{tikz}
\usepackage{siunitx}
\usepackage{stmaryrd}
\usepackage{subcaption}
\usepackage{xcolor}
\usepackage{hyperref}
\usepackage{varwidth}
\usepackage[normalem]{ulem}

\usepgfplotslibrary{external} 
\tikzsetexternalprefix{ext/}

\makeatletter
\g@addto@macro\@floatboxreset\centering
\makeatother

\usetikzlibrary{calc,fadings,shapes.misc,3d,arrows,
                decorations,decorations.pathmorphing,
                decorations.text,shapes.geometric,
                decorations.markings,patterns,spy,positioning,
                arrows.meta,quotes}

\usepackage{tikz-3dplot}
\definecolor{myblue1}{RGB}{0,177,234}
\definecolor{myblue2}{RGB}{76,200,239}
\definecolor{myblue3}{RGB}{127,215,244}
\definecolor{myblue4}{RGB}{178,231,248}
\definecolor{mybluegray1}{RGB}{0,127,167}
\definecolor{mybluegray2}{RGB}{76,165,193}
\definecolor{mybluegray3}{RGB}{127,191,211}
\definecolor{mybluegray4}{RGB}{178,216,228}
\definecolor{mygray1}{RGB}{76,84,93}
\definecolor{mygray2}{RGB}{129,135,141}
\definecolor{mygray3}{RGB}{165,169,174}
\definecolor{mygray4}{RGB}{201,203,206}
\definecolor{myorange1}{RGB}{255,126,46}
\definecolor{myorange2}{RGB}{255,164,108}
\definecolor{myorange3}{RGB}{255,190,150}
\definecolor{myorange4}{RGB}{255,216,192}

\newcommand{\eps}{\varepsilon}

\newcommand{\V}[1]{\textbf{#1}}
\newcommand{\GV}[1]{\boldsymbol{#1}}

\newcommand{\ie}{\textit{i.e.}\;}
\newcommand{\etal}{\textit{et al}\;}

\journal{Computers and Fluids}

\begin{document}

\begin{frontmatter}

\title{U-net architectures for fast prediction of incompressible laminar flows}

\cortext[cor]{~Corresponding author}

\author[cfl]{Junfeng Chen}
\author[cfl]{Jonathan Viquerat\corref{cor}}
\ead{jonathan.viquerat@mines-paristech.fr}
\author[cfl]{Elie Hachem}

\address[cfl]{~MINES~ParisTech, PSL~-~Research~University, CEMEF~-~Centre~for~material~forming,\\ CNRS~UMR~7635, CS~10207~rue~Claude~Daunesse, 06904~Sophia-Antipolis~Cedex, France.}

\begin{abstract}
Machine learning is a popular tool that is being applied to many domains, from computer vision to natural language processing. It is not long ago that its use was extended to physics, but its capabilities remain to be accurately contoured. In this paper, we are interested in the prediction of 2D velocity and pressure fields around arbitrary shapes in laminar flows using supervised neural networks. To this end, a dataset composed of random shapes is built using B\'ezier curves, each shape being labeled with its pressure and velocity fields by solving Navier-Stokes equations using a CFD solver. Then, several U-net architectures are trained on the latter dataset, and their predictive efficiency is assessed on unseen shapes, using \textit{ad hoc} error functions.
\end{abstract}


\end{frontmatter}

\section{Introduction}

During the last few years, the CFD community has largely benefited from the fast-paced development of the machine learning (ML) field, and more specifically for that of the neural networks (NN) domain. In many cases, only a part of the resolution process is replaced with a trained NN, in order to reduce the computational cost. Examples for these applications are the prediction of closure terms in RANS \cite{Ling2016} \cite{Tracey2015} or LES \cite{Beck2018} computations. In other situations, a supervised network is directly trained to predict a flow profile: in \cite{Guo2016}, an auto-encoder NN architecture is used to obtain steady state flow predictions around elementary and real-life shapes; in \cite{Jin2018}, a fusion convolutional neural network (CNN) is trained to predict velocity snapshots around a cylinder in weakly turbulent flows, using the time history of pressure around the cylinder as an input.

\begin{figure}[h!]
\centering

\includegraphics{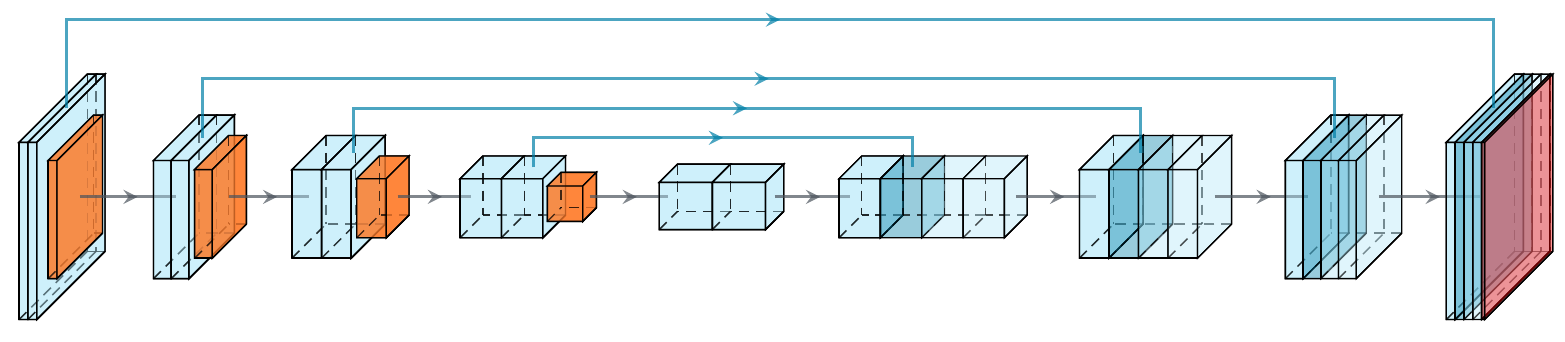}

\caption{\textbf{Standard U-net architecture}. The contractive path is based on a pattern made of two convolutional layers (in light blue) followed by a max-pooling layer (in orange). At each occurence of the pattern, the image size is divided by two, while the number of filters is doubled. In the upscaling path, a deconvolution step is first applied to the input, while the number of filters is halved. The output of this layer is then concatenated with its mirror counterpart in the contractive path using a shortcut connections (here noted with arrows). Finally, two convolution layers are applied. At the end of the last layer, a $1 \times 1$ convolution with 3 filters is applied to obtain an RGB image at the output (in red).}
\label{fig:Unet}
\end{figure}

The U-net is a particular CNN architecture, which was initially proposed by Ronneberger \etal \cite{Ronneberger2015} for biomedical image segmentation. As shown in figure \ref{fig:Unet}, U-nets are composed of (i) a contractive path based on the repetition of a convolution/max-pooling pattern, followed by (ii) an upscaling path, in which a pattern of deconvolution/concatenation/convolution is applied several times. The \emph{skip connections} in U-nets, originally proposed by \cite{Long2015}, are used to link the contracting path to the upscaling path, and help localize features of high resolution during the upsampling, thus improving the segmentation accuracy.

Based on the end-to-end property of U-nets, several variations have been proposed in the recent years. In \cite{Newell2016}, a stacked network architecture is proposed for the estimation human pose. The whole network is simply built by adding U-nets one after the other, the output of a U-net being the input of the following one. In \cite{Tang2018}, shortcuts were added from each U-net to all the following ones, in order to help the back-propagation of gradient. Multi-paths U-nets were introduced in \cite{Dolz2019}, where separated contracting paths were used for images holding different informations. The different paths are merged at the bottom of the U-net, and share a single expanding path. In \cite{Li2019}, the authors report the parallel training of independent U-nets, the final prediction being obtained as a weighted sum of the outputs of the different networks. Other methodological refinements demonstrated their potential, such as multi-scale inputs and supervision \cite{Stawiaski2017}, for example. These different variations are discussed in details in section \ref{section:archis}.

U-net architectures recently made their way to other physical fields, such as fluid dynamics. In \cite{Thuerey2018}, the authors exploit U-nets to infer the velocity and pressure of turbulent flow around airfoils computed in a Reynolds-Averaged Navier-Stokes framework. In \cite{Fukami2018}, a traditional CNN is coupled to a Downsampled Skip-Connection Multi-Scale (DSC/MS) model to reconstruct the turbulent flow computed by direct numerical simulation. The DSC/MS model, which uses a downscaling-upsampling path like U-net, efficiently recovers the detailed velocity field. 

In this paper, focus is made on the prediction of laminar flows around arbitrary 2D shapes. To this end, we assess the predictive capabilities of several U-net based architectures, and compare them to that of a 4-levels baseline U-net, presented in figure \ref{fig:Unet}. In section \ref{section:archis}, the functioning of the baseline U-net is briefly covered, and advanced architectures are introduced. Details about the dataset are given in section \ref{section:dataset}, along with a discussion on the metrics used to evaluate the results of the different architectures. In section \ref{section:training}, details are given about the training, and the performances of the different networks are compared on a test subset from the dataset. Finally, the predictions of the networks are assessed on unseen configurations, such as geometrical shapes or airfoils.

\section{Existing U-net architectures}
\label{section:archis}

U-nets were introduced in 2015 \cite{Ronneberger2015} as an improvement over a sliding-window method for classical convolutional networks \cite{Ciresan2012}. In classical convolutional networks, a contracting path similar to the first branch of the U-net is used to extract high-level features from an input image. However, as the abstraction level of the features increases, localization information about these features is lost, making it hard to directly reconstruct meaningful image as an output of the network. The idea behind U-net is to follow the contracting path by an upscaling path, in which the resolution of the feature maps is re-increased symmetrically to the contracting path using deconvolution layers. After each deconvolution, localization information is re-injected in the upscaling path by concatenating information from the symmetric level of the downscaling path using a shortcut connection. The concatenation is followed by a convolution operation that assembles the two informations (features coming from the previous level of the upscaling path, and localization coming from the shortcut connection) together.

\begin{figure}
\centering

\begin{subfigure}[b]{.4\textwidth}
	\centering
	\includegraphics{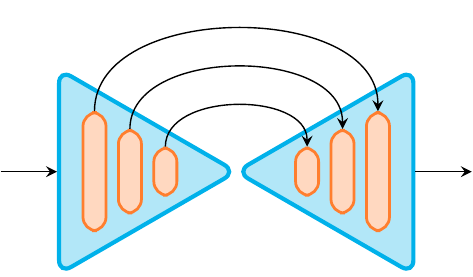}
	\caption{\label{fig:Unets}Standard U-net}
\end{subfigure} \quad
\begin{subfigure}[b]{.55\textwidth}
	\centering
	\includegraphics{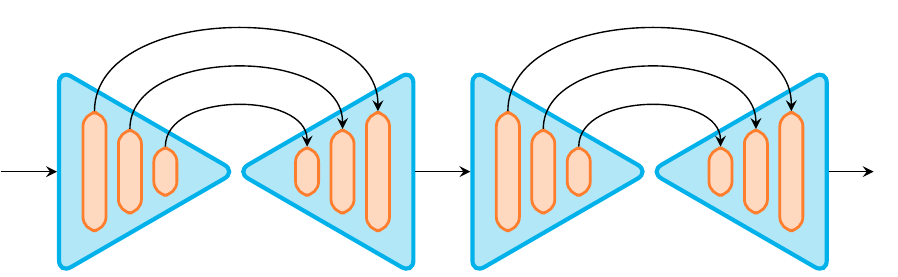}
	\caption{\label{fig:SUnets}Stacked U-nets}
\end{subfigure}

\begin{subfigure}[b]{.4\textwidth}
	\centering
	\includegraphics{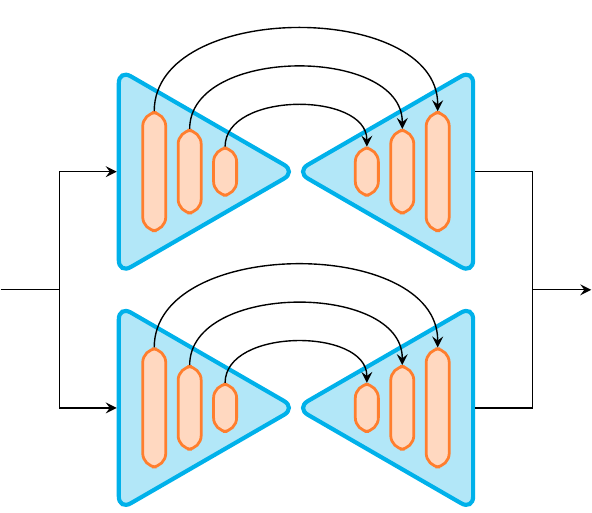}
	\caption{\label{fig:PUnets}Parallel U-nets}
\end{subfigure} \quad
\begin{subfigure}[b]{.55\textwidth}
	\centering
	\includegraphics{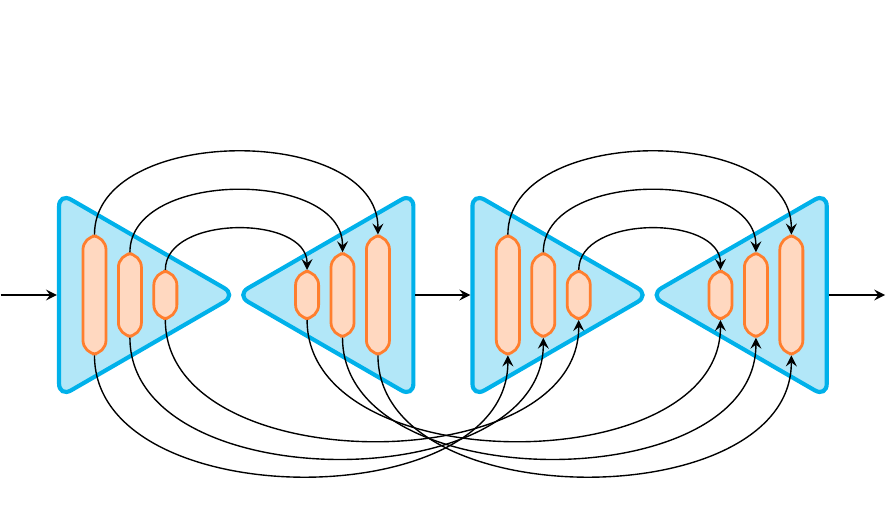}
	\caption{\label{fig:CUnets}Coupled U-nets}
\end{subfigure}
\caption{\textbf{U-net architectures}.}
\label{fig:archis}
\end{figure}

Recently, improved architectures based on U-nets have been proposed. Newell \etal \cite{Newell2016} proposed a stacked architecture, similar to the one shown in figure \ref{fig:SUnets}. The successive stacking of networks is said to consolidate the captured features, and to allow for the grasp of more complex spatial relationships. In \cite{Tang2018}, Tang \etal introduced coupled U-nets (CU-nets), presented in figure \ref{fig:CUnets}. Based on the current trend of dense connectivities in convolutional networks \cite{Huang2017}, the authors proposed to add a similar level of density to coupled U-nets, by introducing shortcuts between layers from same levels. They show that these shortcut help the information flow across the structure, while improving the efficiency of the gradient backpropagation. Finally, the separate training of U-nets in parallel is reported in \cite{Li2019}. In this contribution, the authors generate different versions of their initial dataset by adding different zero-mean gaussian noises. Then, three different U-nets are trained separately, each on a different version of the dataset. The final output is computed as a weighted combination of the outputs of the separate U-nets. Results show that this method reduces overfitting, thus inducing better performances on reduced datasets, and proved to be superior to a single U-net trained on any of the datasets (noisy or not).

In the current paper, we introduce a parallel U-net (PU-net) architecture, shown in figure \ref{fig:PUnets}. The input image is fed to two separate U-nets, which outputs are simply recombined using an average layer. At the difference of \cite{Li2019}, the two networks are trained concurrently, using the same dataset.

\section{Dataset construction}
\label{section:dataset}

This section provides insights on the random shape dataset generation used to train networks in the next sections. This dataset was initially used in \cite{Viquerat2019}, thus only the main lines are sketched here. For more details, the reader is referred to \cite{Viquerat2019}. First, we describe the steps to generate arbitrary shapes by means of connected Bezier curves. Then, the solving of the Navier-Stokes equations with an immersed method is presented. Finally, details about the dataset are given.

\subsection{Random shape generation}

\begin{figure}
\centering

\begin{subfigure}{.45\textwidth}
	\centering
	\includegraphics{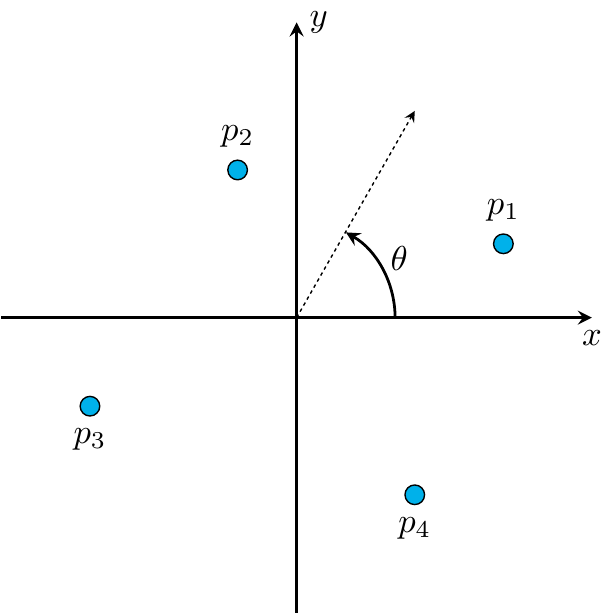}
	\caption{Draw $n_s$ random points, translate them around $(0,0)$ and sort them by ascending trigonometric angle}
	\label{fig:shape_generation_1}
\end{subfigure} \qquad
\begin{subfigure}{.45\textwidth}
	\centering
	\includegraphics{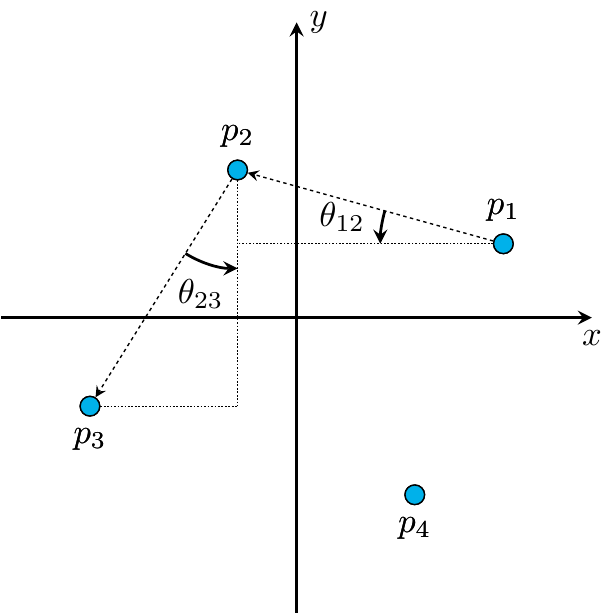}
	\caption{Compute angles between random points, and compute an average angle around each point $\theta^*_i$}
	\label{fig:shape_generation_2}
\end{subfigure}

\medskip
\medskip

\begin{subfigure}{.45\textwidth}
	\centering
	\includegraphics{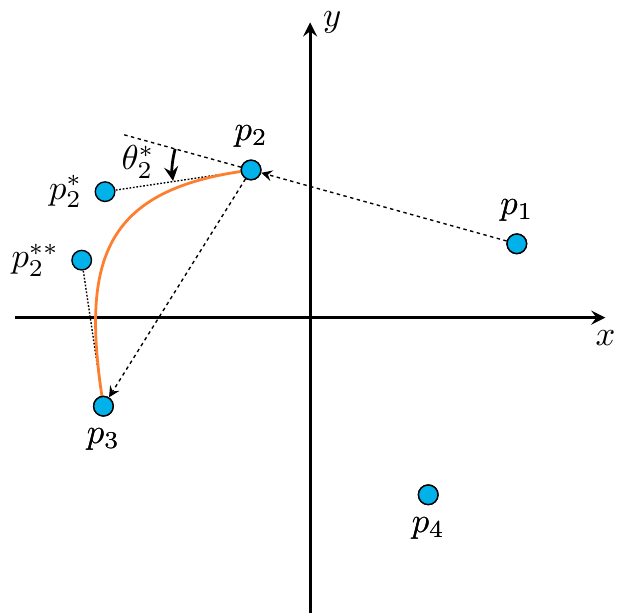}
	\caption{Compute control points coordinates from averaged angles and generate cubic B\'ezier curve}
	\label{fig:shape_generation_3}
\end{subfigure} \qquad
\begin{subfigure}{.45\textwidth}
	\centering
	\includegraphics{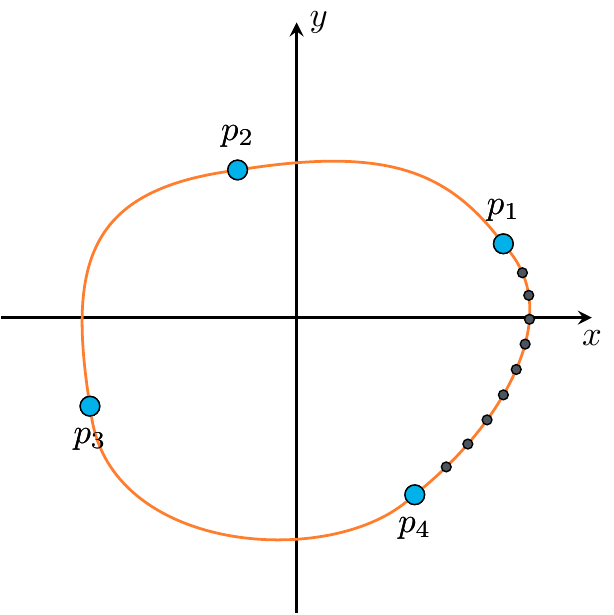}
	\caption{Sample all B\'ezier lines and export for mesh immersion}
	\label{fig:shape_generation_4}
\end{subfigure}
\caption{\textbf{Random shape generation with cubic B\'ezier curves}. }
\label{fig:shape_generation}
\end{figure}

\begin{figure}[h!]
\centering

\def\w{1.5cm}
\def\arraystretch{4.5}
\setlength\tabcolsep{9pt}
\newcolumntype{C}{ >{\centering\arraybackslash} m{1.5cm} }

\makebox[\linewidth]{%
	\begin{tabular}{CCCCCCC}
		\includegraphics[width=\w]{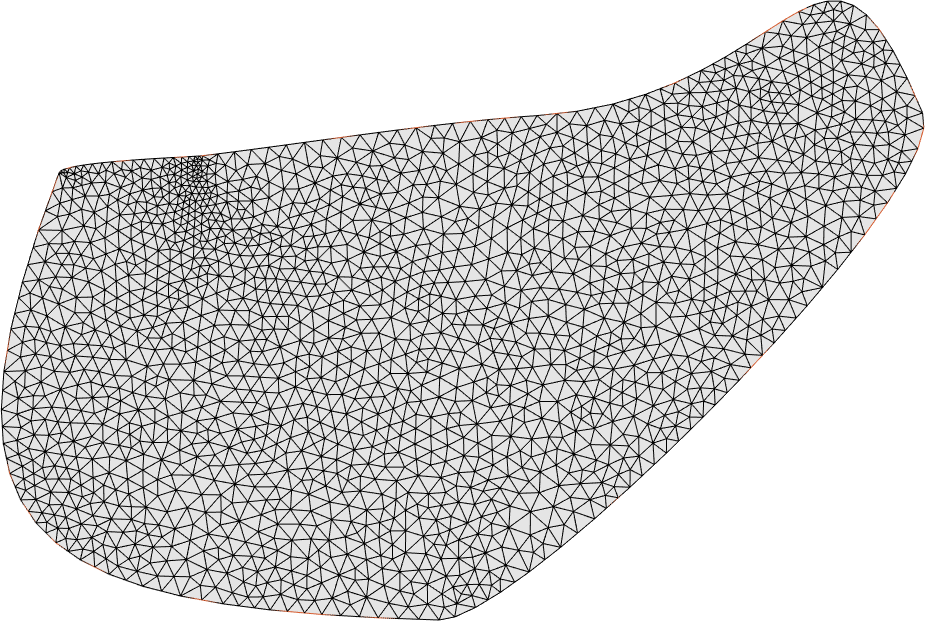} 		& \includegraphics[width=\w]{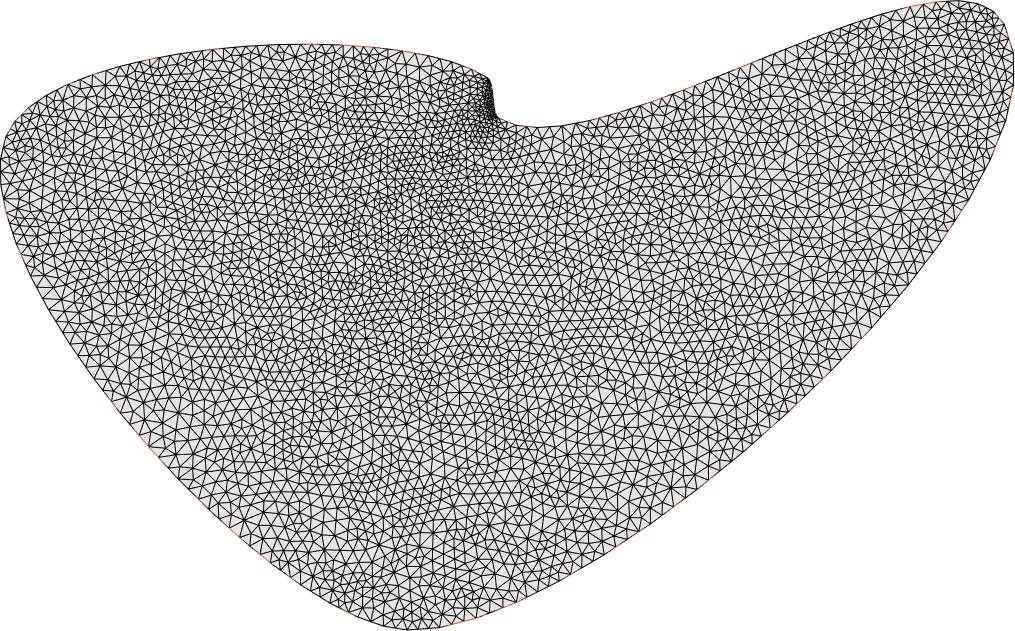}
											& \includegraphics[width=\w]{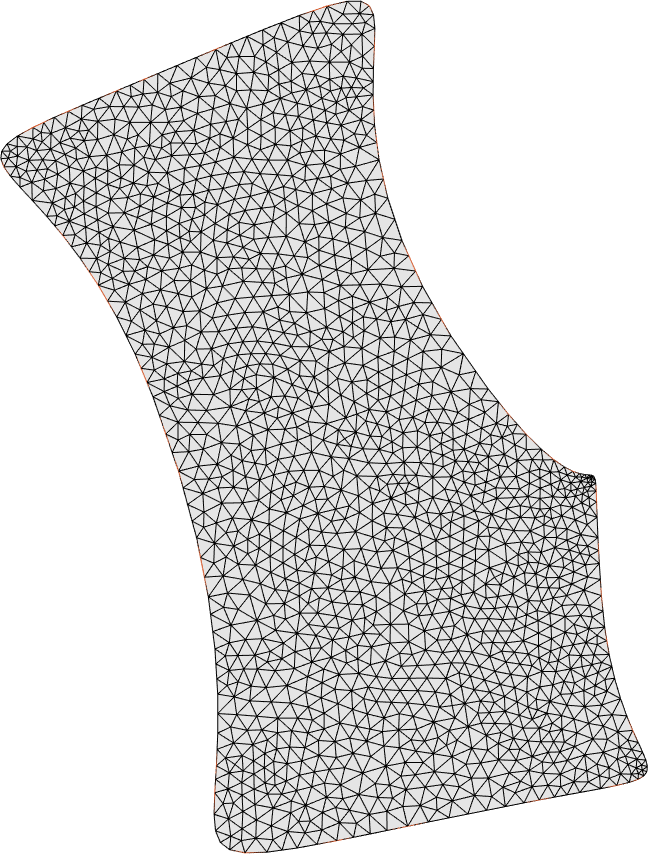}
											& \includegraphics[width=\w]{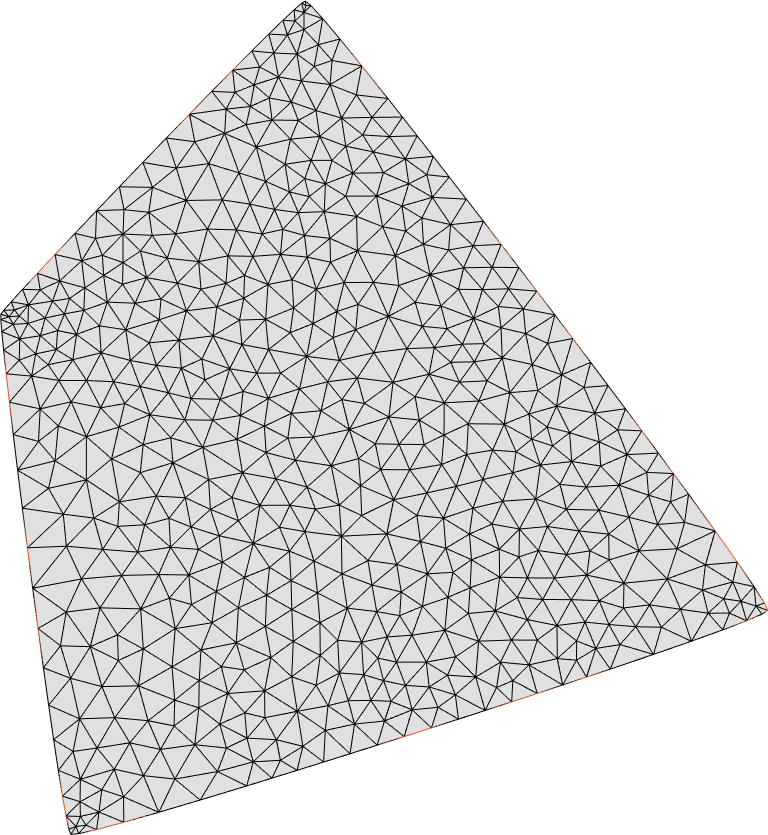}
											& \includegraphics[width=\w]{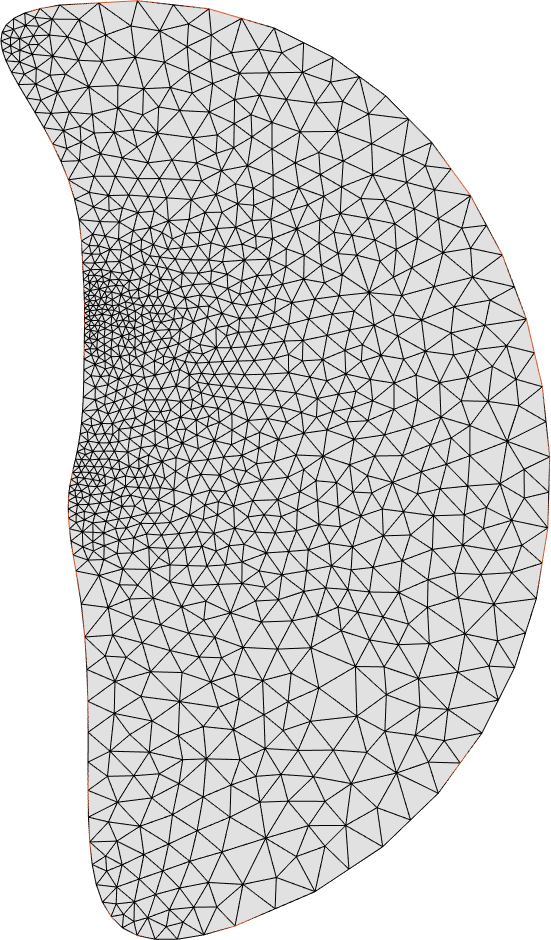}
											& \includegraphics[width=\w]{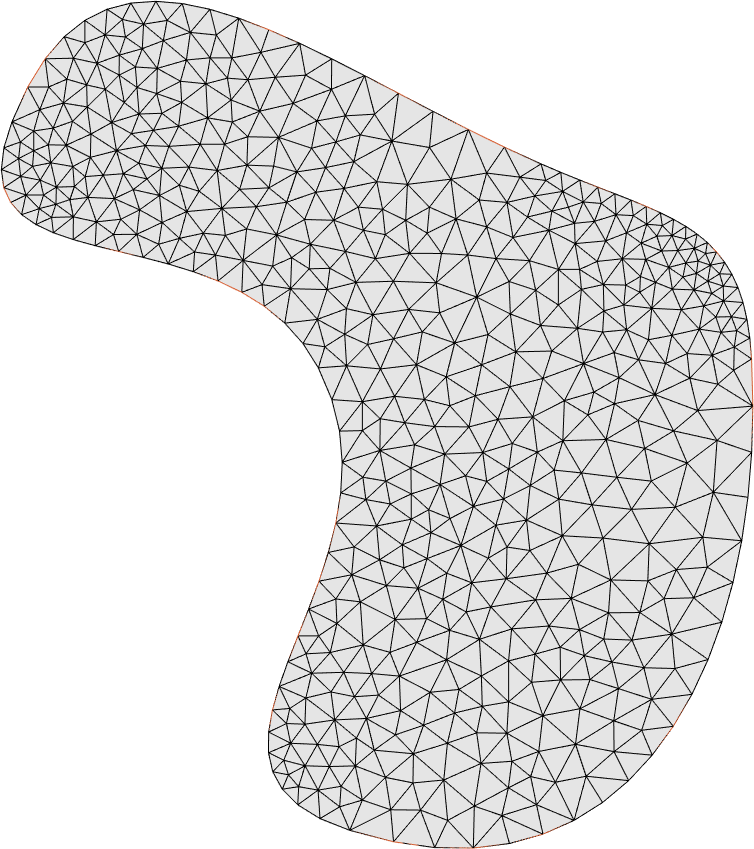}
											& \includegraphics[width=\w]{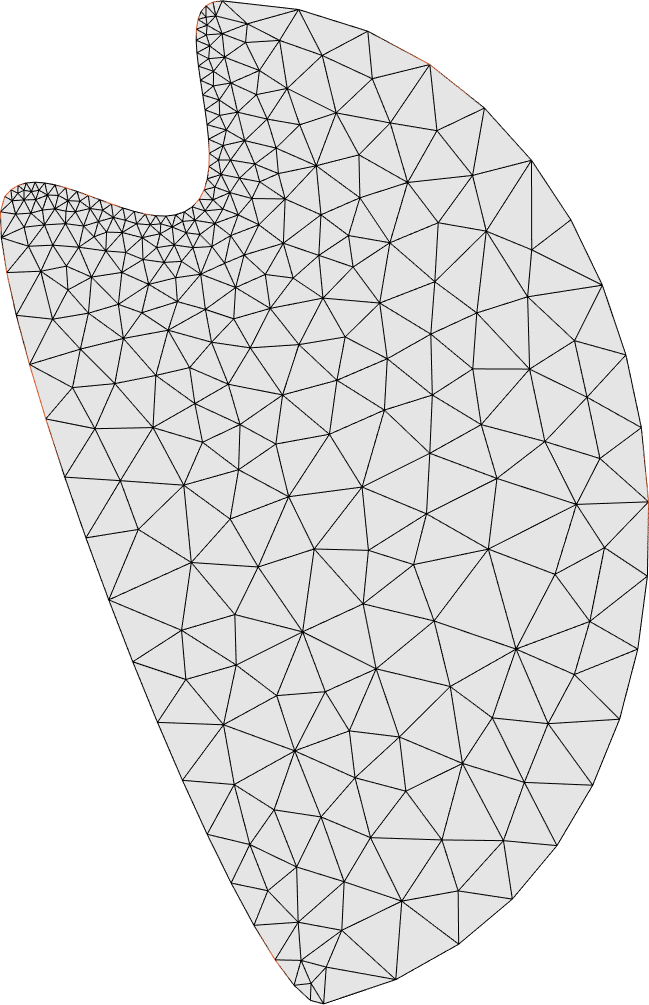}\\
		\includegraphics[width=\w]{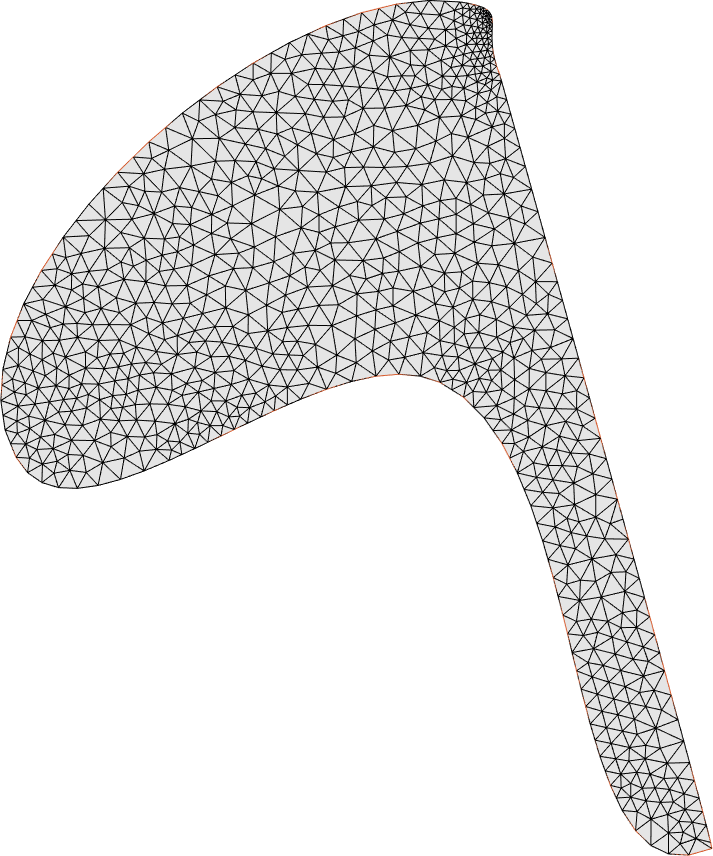} 		& \includegraphics[width=\w]{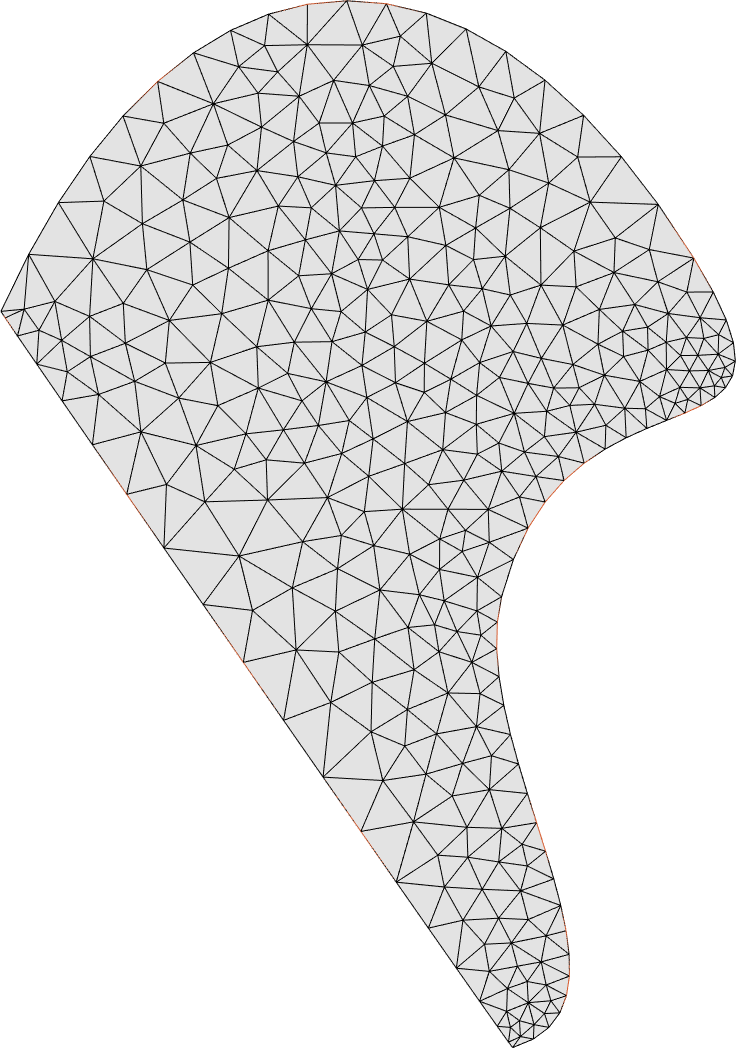}
											& \includegraphics[width=\w]{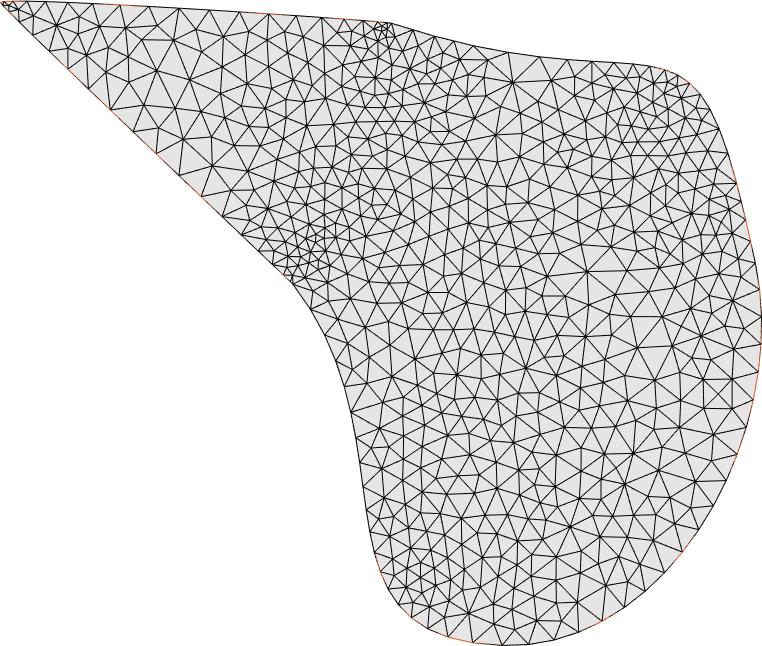}
											& \includegraphics[width=\w]{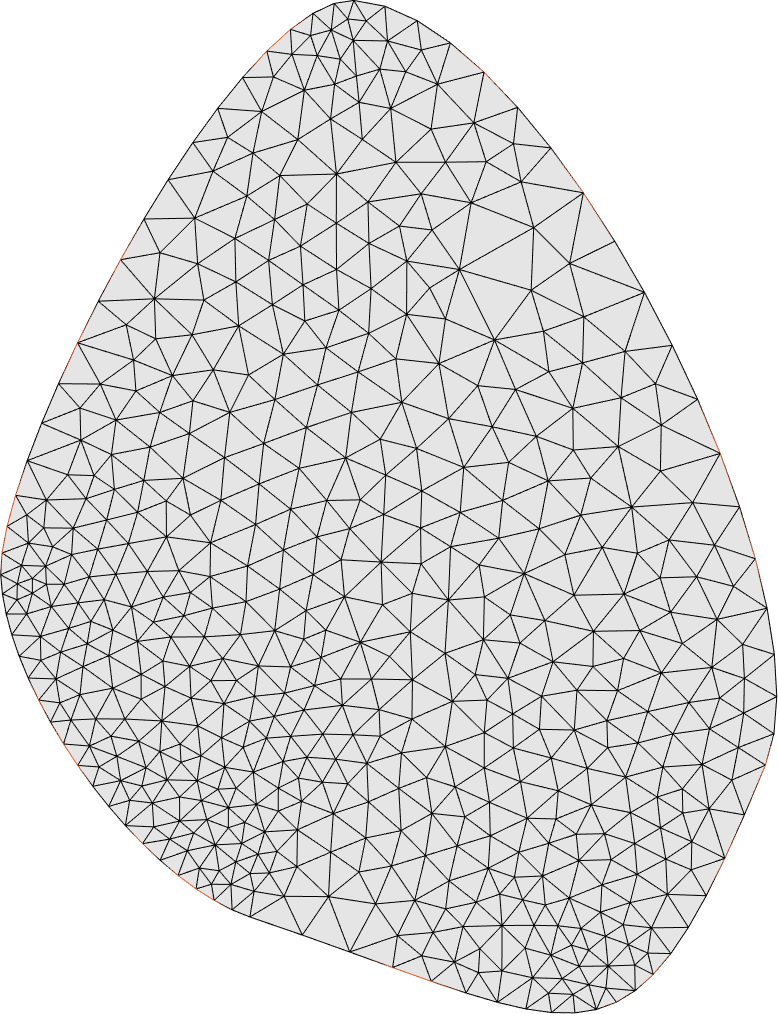}
											& \includegraphics[width=\w]{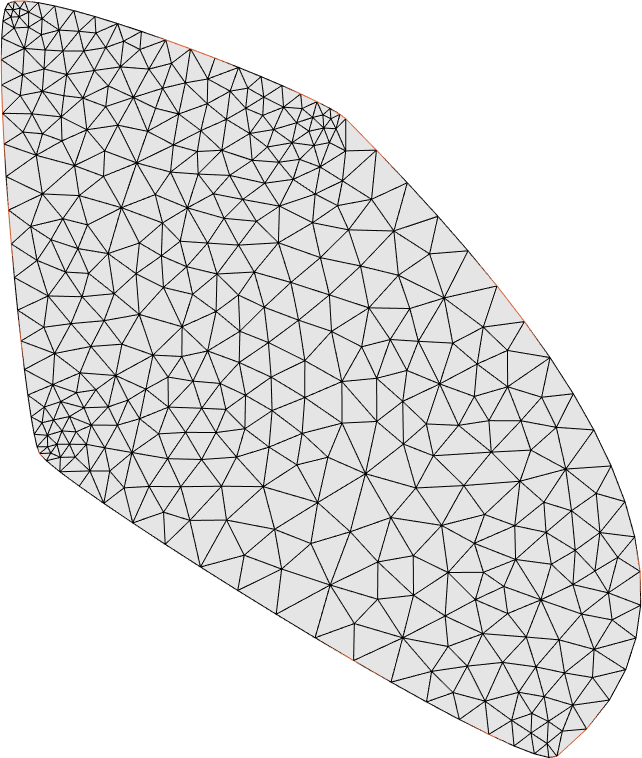}
											& \includegraphics[width=\w]{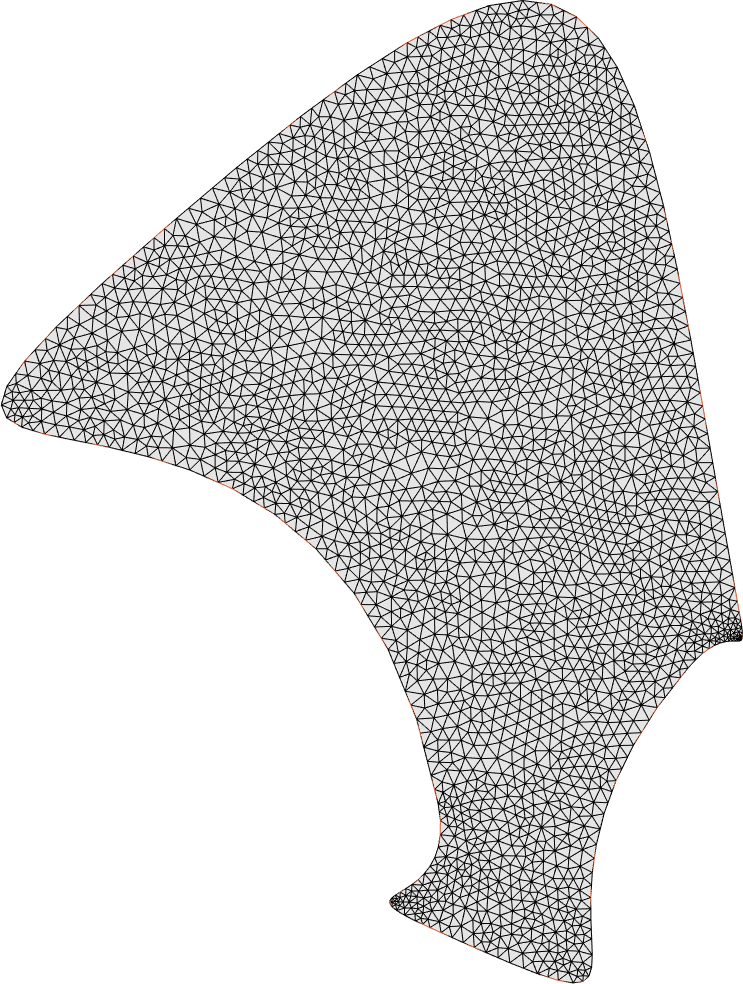}
											& \includegraphics[width=\w]{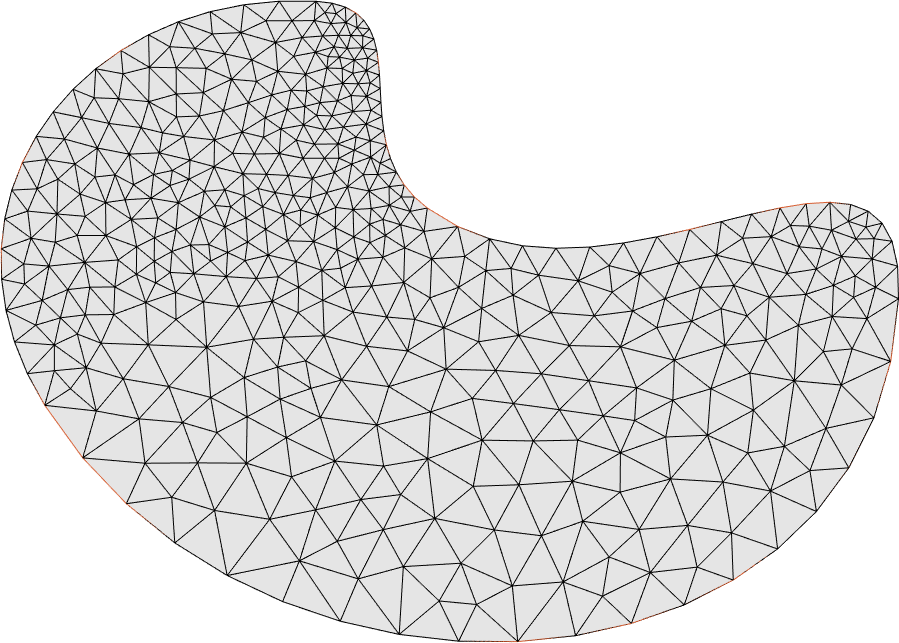}\\
		\includegraphics[width=\w]{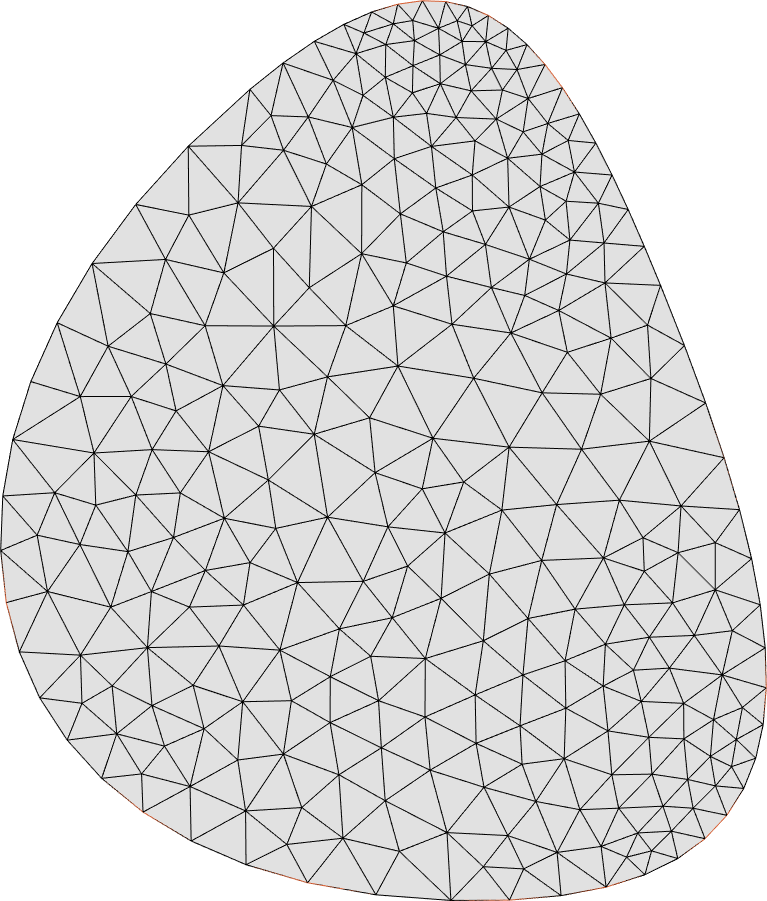} 	& \includegraphics[width=\w]{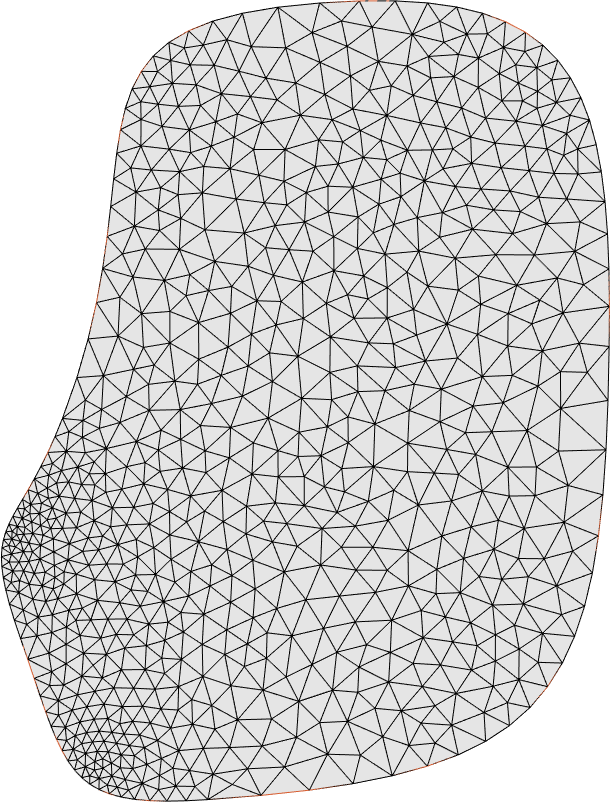}
											& \includegraphics[width=\w]{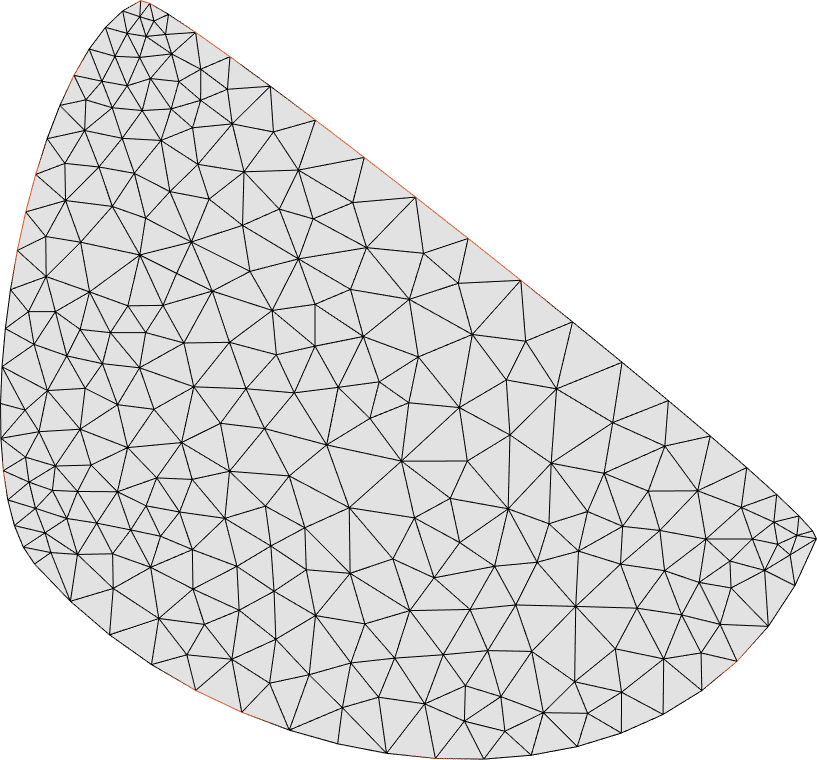}
											& \includegraphics[width=\w]{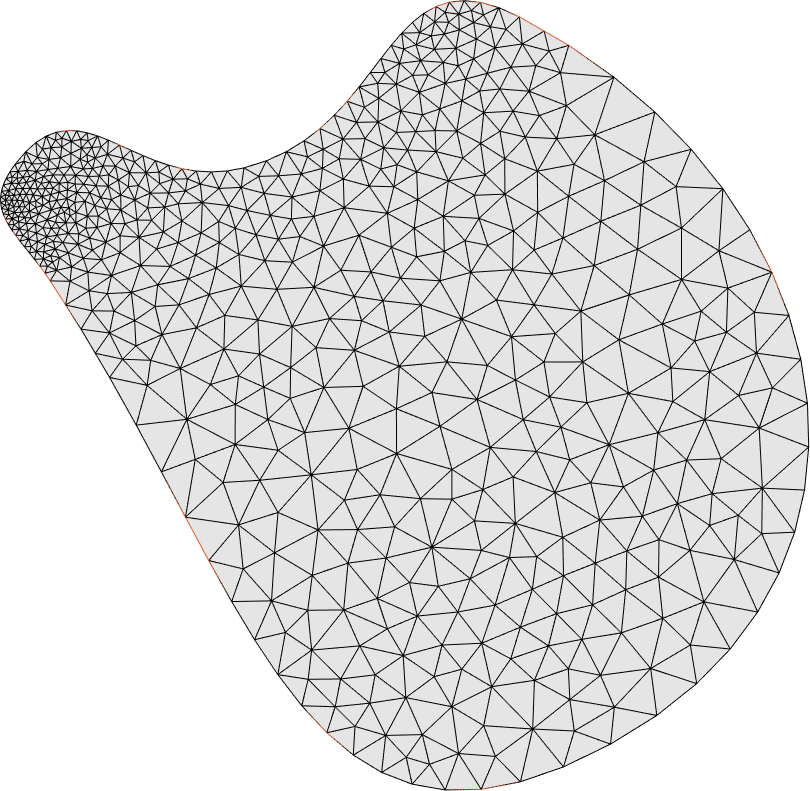}
											& \includegraphics[width=\w]{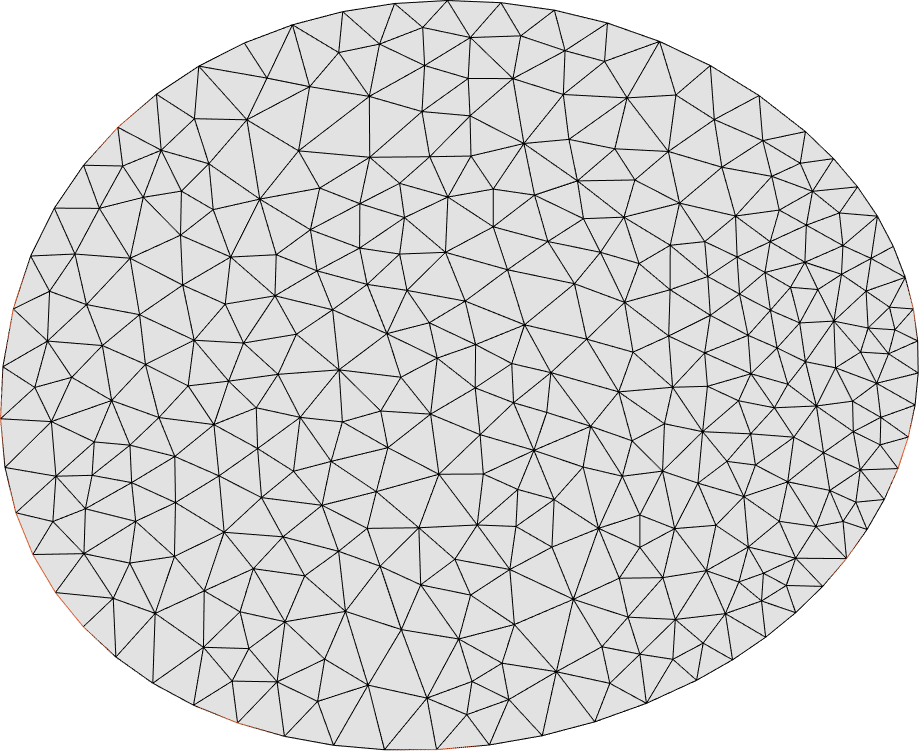}
											& \includegraphics[width=\w]{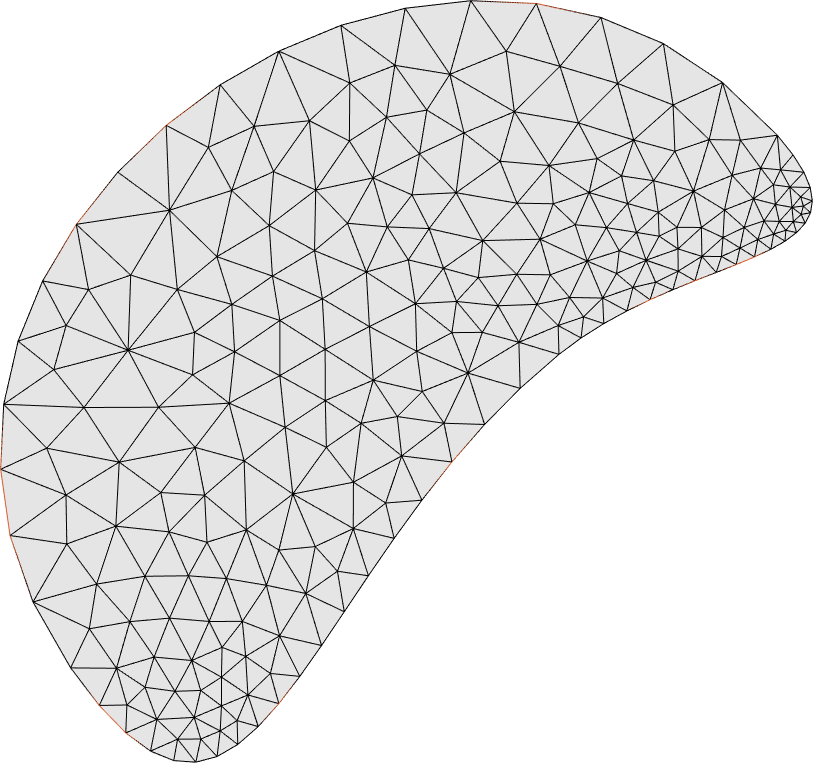}
											& \includegraphics[width=\w]{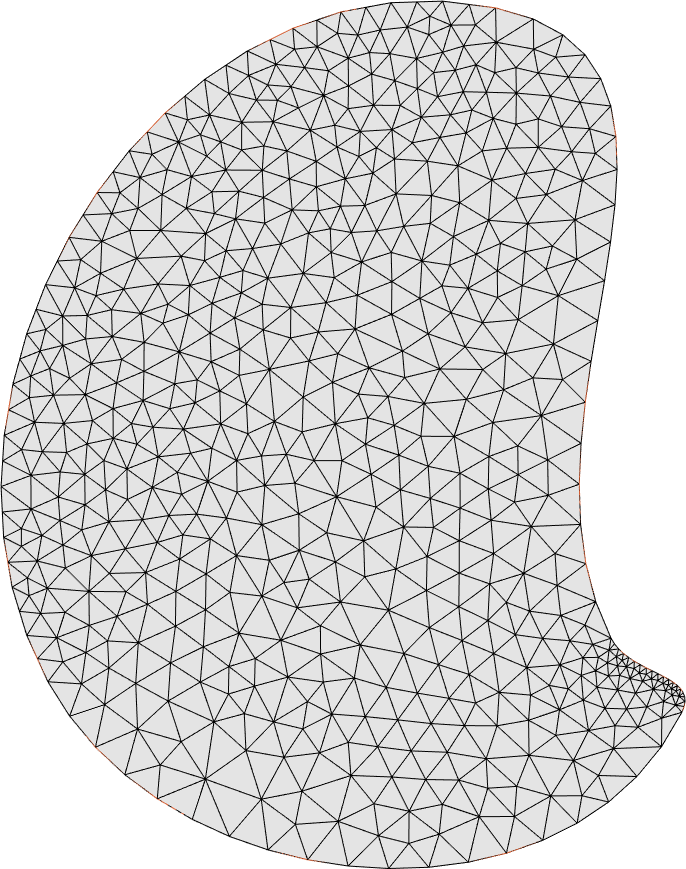}%
\end{tabular}}

\caption{\textbf{Shape examples drawn from the dataset.} A wide variety of shape is obtained using a restrained number of points ($n_s \in \left[ 4, 6 \right]$), as well as a local curvature $r$ and averaging parameter $\alpha$.}
\label{fig:shape_examples}
\end{figure}

In the first step, $n_s$ random points are drawn in $\left[ 0, 1 \right]^2$, and translated so their center of mass is positioned in $(0,0)$. An ascending trigonometric angle sort is then performed (see figure \ref{fig:shape_generation_1}), and the angles between consecutive random points are then computed. An average angle is then computed around each point (see figure \ref{fig:shape_generation_2}) using:

\begin{equation*}
	\theta^*_i = \alpha \theta_{i-1,i} + (1 - \alpha) \theta_{i,i+1},
\end{equation*}

\noindent with $\alpha \in \left[0,1\right]$. The averaging parameter $\alpha$ allows to alter the sharpness of the curve locally, maximum smoothness being obtained for $\alpha = 0.5$. Then, each pair of points is joined using a cubic B\'ezier curve, defined by four points: the first and last points, $p_i$ and $p_{i+1}$, are part of the curve, while the second and third ones, $p^*_i$ and $p^{**}_i$, are control points that define the tangent of the curve at $p_i$ and $p_{i+1}$. The tangents at $p_i$ and $p_{i+1}$ are respectively controlled by $\theta^*_i$ and $\theta^*_{i+1}$ (see figure \ref{fig:shape_generation_3}). A final sampling of the successive B\'ezier curves leads to a boundary description of the shape (figure \ref{fig:shape_generation_4}). Using this method, a wide variety of shapes can be attained, as shown in figure \ref{fig:shape_examples}.

\subsection{Numerical resolution of Navier-Stokes equations}
\label{section:NS}

The flow motion of incompressible newtonian fluids is described by the Navier-Stokes (NS) equations: 

\begin{equation} \label{eq:ns_equation1}
	\left\{
	\begin{aligned}
		\rho\ (\partial_{t} \V{v} + \V{v} \cdot \nabla \V{v}) -\nabla \cdot \left( 2 \eta \GV{\epsilon}(\V{v}) - p \V{I} \right)  & = \V{f}, \\
		\nabla \cdot \V{v} &= 0,
	\end{aligned}
	\right.
\end{equation}

\noindent where $t \in [0,T]$ is the time, $\V{v}(x,t)$ the velocity, $p(x,t)$ the pressure, $\rho$ the fluid density, $\eta$ the dynamic viscosity and $\V{I}$ the identity tensor. In order to efficiently construct the dataset, an immersed boundary method is used for resolution instead of the usual body-fitted method, avoiding a systematic re-meshing of the whole domain for each shape. This method rely on a unified fluid-solid eulerian formulation based on level-set description of the geometry \cite{Bruchon2009}, and leads to the following set of modified equations: 

\begin{equation} \label{eq:ns_equation2}
	\left\{
	\begin{aligned}
		\rho^* (\partial_{t} \V{v} + \V{v} \cdot \nabla \V{v}) -\nabla \cdot \left( 2 \eta \GV{\epsilon}(\V{v}) + \GV{\tau} - p \V{I} \right)  & = \V{f}, \\
		\nabla \cdot \V{v} &= 0,
	\end{aligned}
	\right.
\end{equation}

\noindent where we have introduced the following mixed quantities:

\begin{equation*}
	\begin{aligned}
		\GV{\tau} & = H(\alpha) \GV{\tau}_{\text{s}},\\
		\rho^* & = H(\alpha) \rho_{\text{s}} + (1-H(\alpha)) \rho_{\text{f}},
	\end{aligned}
\end{equation*}

\noindent where the subscripts $f$ and $s$ respectively refer to the fluid and the solid, and $H(\alpha)$ is the Heaviside function:

\begin{equation} \label{eq:heavyside31}
	H(\alpha) = \left\{
	\begin{aligned}
		1 & \text{ if}\ \alpha > 0,\\
		0 & \text{ if}\ \alpha < 0.
	\end{aligned}
	\right.
\end{equation}

\noindent The reader is referred to \cite{Hachem2013} for additional details about formulation (\ref{eq:ns_equation2}). Eventually, the modified equations (\ref{eq:ns_equation2}) are cast into a stabilized finite element formulation, and solved using a variational multi-scale (VMS) solver \cite{Hachem2013}.

\subsection{Dataset}

\begin{figure}[h!]
\centering

\setlength{\fboxsep}{0pt}%
\setlength{\fboxrule}{1pt}%

\def\scale{0.738}

\begin{subfigure}{.3\textwidth}
	\centering
	\fbox{\includegraphics[width=\linewidth]{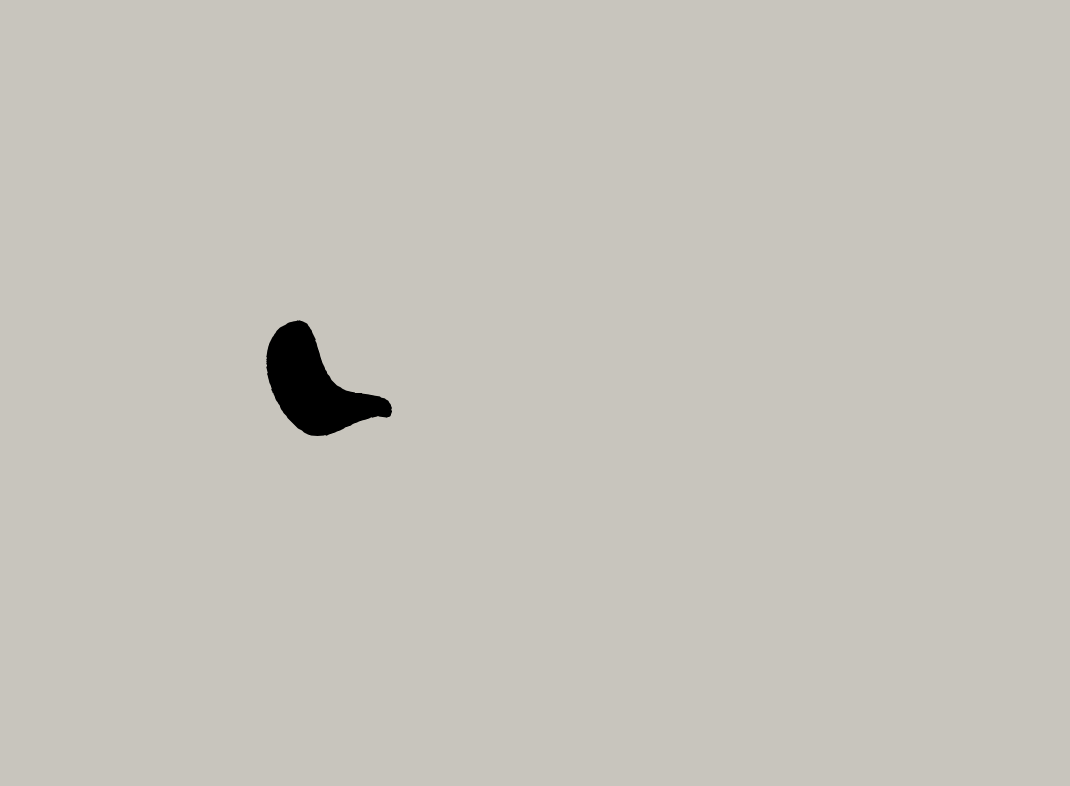}} 
	\caption{Network input}
\end{subfigure} \quad
\begin{subfigure}{.3\textwidth}
	\centering
	\fbox{\includegraphics[width=\linewidth]{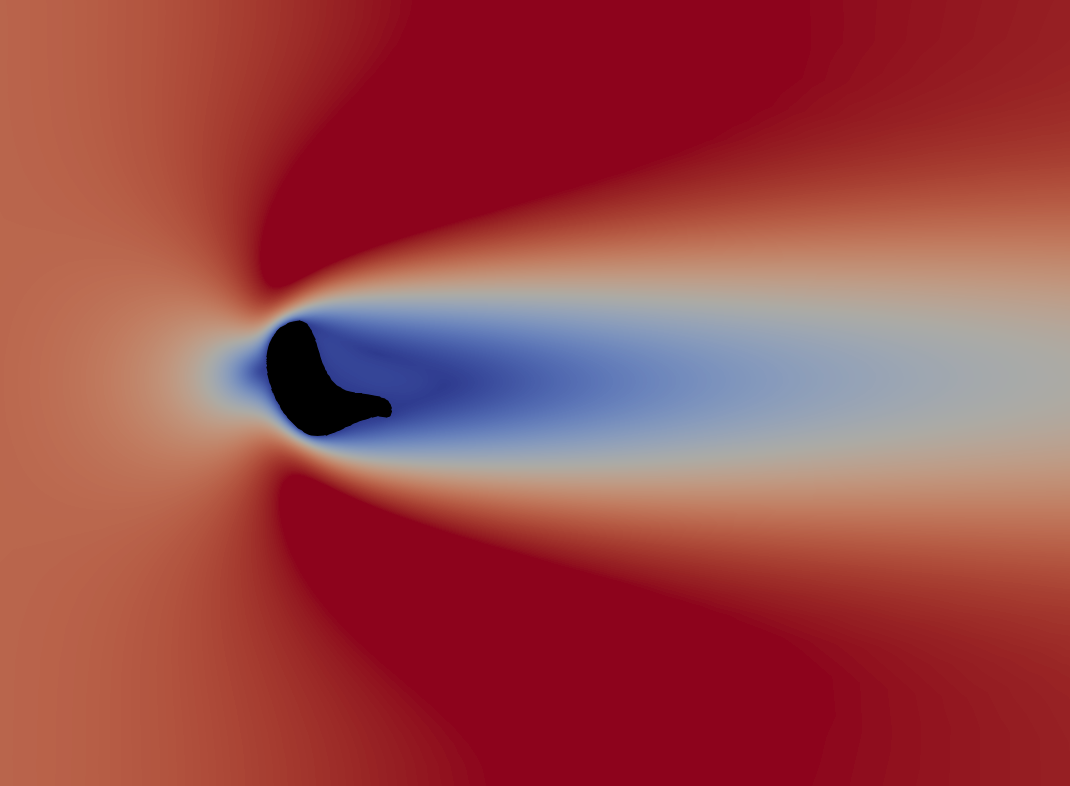}}
	\caption{Velocity field}
\end{subfigure} \quad
\begin{subfigure}{.3\textwidth}
	\centering
	\fbox{\includegraphics[width=\linewidth]{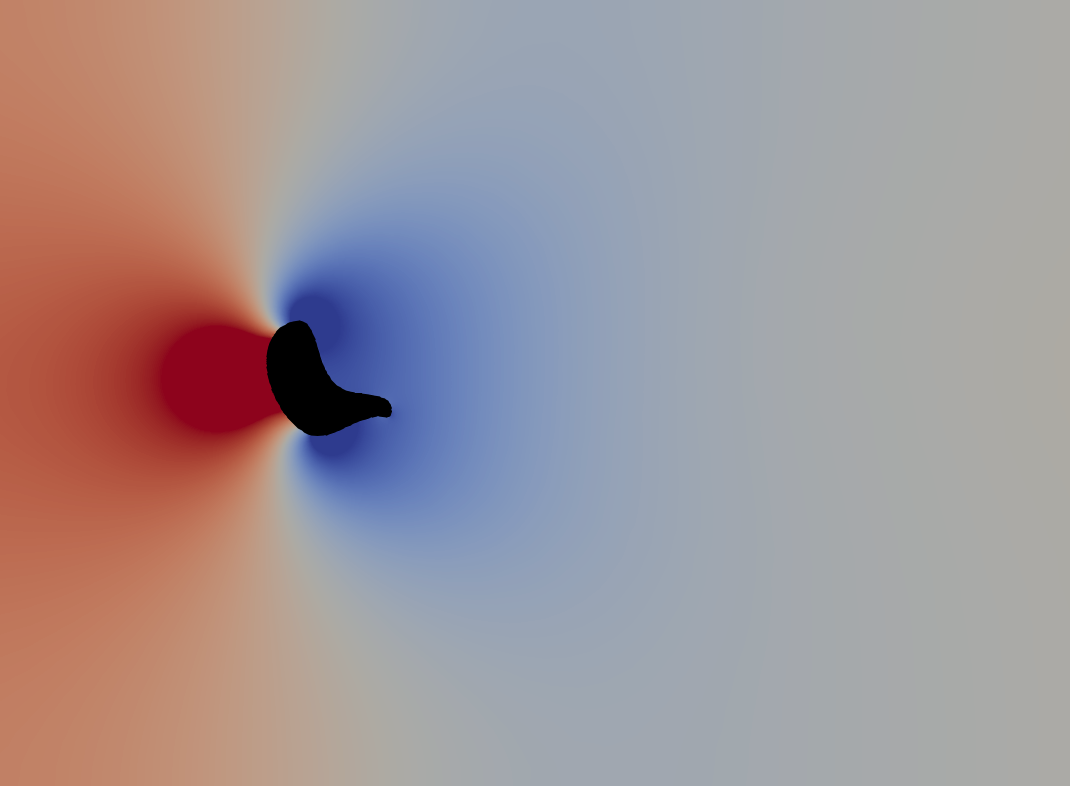}}
	\caption{Pressure field}
\end{subfigure}

\caption{\textbf{Network input, velocity field and pressure field for a dataset element.}. The shape is shown in its computational domain (left), along with the computed velocity field (center) and pressure field (right). Fields are scaled to $\left[ 0, 1 \right]$ for velocity and $\left[ -1, 1 \right]$ for pressure.}
\label{fig:dataset_example}
\end{figure}

The dataset (DS) is composed of 12.000 shapes, along with their steady-state velocity and pressure fields at $Re=10$ (see figure \ref{fig:dataset_example}). All the labels were computed using cimLib \cite{Hachem2013}, following the methods exposed in section \ref{section:NS}. In the following, the DS is systematically divided into three sets: 9600 shapes for the training set, 1200 shapes for the validation set, and 1200 shapes for the test set.

\section{Networks training}
\label{section:training}

\subsection{Networks}

The baseline U-net exploited in this paper is composed of 4 blocks in the contractive path, a bottleneck block, and 4 blocks in the upscaling path (see figure \ref{fig:Unet}. The initial number of kernels is set to 32 and is doubled at each pooling operation, reaching 512 in the bottleneck. It is then divided by two at each upscaling step, dropping back to 32 in the final block. The upscaling path is followed by a last $1 \times 1$ convolution with 3 filters, thus reconstructing an RGB image. All convolution kernels are of size $3 \times 3$, deconvolution kernels and strides of size $2 \times 2$, and pooling sizes and strides also of size $2 \times 2$. All activation functions are rectified linear units (ReLu).

\subsection{Training}

The learning process in neural networks consists in adjusting all the biases and weights of the network in order to reduce the value of a well-chosen loss function. For regression cases, it is common to choose the mean squared error (MSE). With the loss function at hand, the optimization of the weights and biases is performed with a stochastic gradient descent that exploits the back-propagation algorithm \cite{Goodfellow2017}. The gradient descent is usually performed multiple times on small random subsets (called mini-batches), instead of considering a single evaluation of the gradient on the whole dataset. This method represents a compromise between an accurate gradient computation and a low computational cost. In this paper, the batch size is set to 8. The whole dataset, split in mini-batches, is re-used multiple times in random orders, a full re-use of it being called an epoch. This procedure is repeated as long as the validation accuracy (the accuracy of the prediction computed on the validation subset) increases, and is stopped when it starts decreasing, meaning that the network is overfitting (\ie learning non-generic features) on the training subset. In this paper, the input images are downscaled to a constant size of $107 \times 78$ pixels to limit the training cost. To ensure fair comparisons, all random seeds are set once and for all for the training of the different considered networks.

The amount of ready-to-use neural networks libraries has exploded in the recent years, most of them exploiting C++ or Python. For supervised learning, they usually include a wide range of choices regarding layer types, activation functions, losses, optimizers and so on. In this paper, we chose to use Keras \cite{Chollet2018} for its high level of abstraction and the ease of use provided by the Python language.

\subsection{Model evaluation}
\label{section:evaluation}

The mean squared error is often used as a loss function for the train of neural networks. It is defined as:

\begin{equation}
	\eps_{\text{mse}} = \frac{1}{n(P)} \sum_{p=1}^{n(P)} \left( \frac{1}{3} \sum_{j=1}^3 \left(y_{pj} - \hat{y}_{pj}\right)^2 \right),
\end{equation}

\noindent where $P$ is the set of all pixels in an image, $n(P)$ is the number of pixels in the image, $j$ is the index looping over the three channels of an RGB image, $\hat{y}_{pj}$ is the predicted value of channel $j$ at pixel $p$, and $y_{pj}$ is its corresponding label. Still, it can be hard to compare the quality of different predictions by relying only on the mean squared error, as it is directly related to the amplitude of the solution. To this end, a complementary pixel-wise relative error function is proposed for the assessment of the predictions quality:

\begin{equation}
	\eps_{\alpha} = \frac{1}{n(P)} \sum_{p=1}^{n(P)} \mathbbm{1} \left( \eps^p_\alpha > 0 \right), \text{ with } \eps^p_\alpha = \mathbbm{1} \left( \left| \frac{y_{p} - \hat{y}_{p}}{y_{p} + \eta} \right| > \alpha \right) \text{ and } y_p = \sum_{j=1}^3 y_{pj}.
\end{equation}

\noindent In the latter expression, $\eta$ is a small positive number introduced to ensure a non-zero denominator in the case where $y_{pj} = 0$, and $\alpha$ is a tolerance for the relative error. Hence, this metric returns the proportion of pixels which prediction accuracy is lower than a given threshold level $\alpha$, without being proportional to the solution amplitude. Regarding the perturbation induced by $\eta$, we argue that only the pixels whose absolute value is close to $\eta$ will be impacted. By setting $\eta = \num{1e-6}$, the variations observed in $\eps_\alpha$ were systematically lower than $0.1\%$, which represents a negligible error, as will be shown in later sections. Additionally, for a single image, it is possible to plot the 2D map of $\eps^p_\alpha$ to locate the areas of low prediction accuracy.

\section{Results}

In this section, a prediction from the test subset obtained with the baseline U-net is first presented and analyzed in details with the two metrics presented in section \ref{section:evaluation}. Then, the different U-net architectures are statistically compared on the test subset. In a third time, the predictive efficiency of the different networks is assessed on several unseen shapes, including NACA airfoils.

\begin{figure}
\centering

\setlength{\fboxsep}{0pt}%
\setlength{\fboxrule}{1pt}%

\def\scale{0.55}

\begin{subfigure}{.45\textwidth}
	\centering
	\fbox{\includegraphics[width=\scale\linewidth]{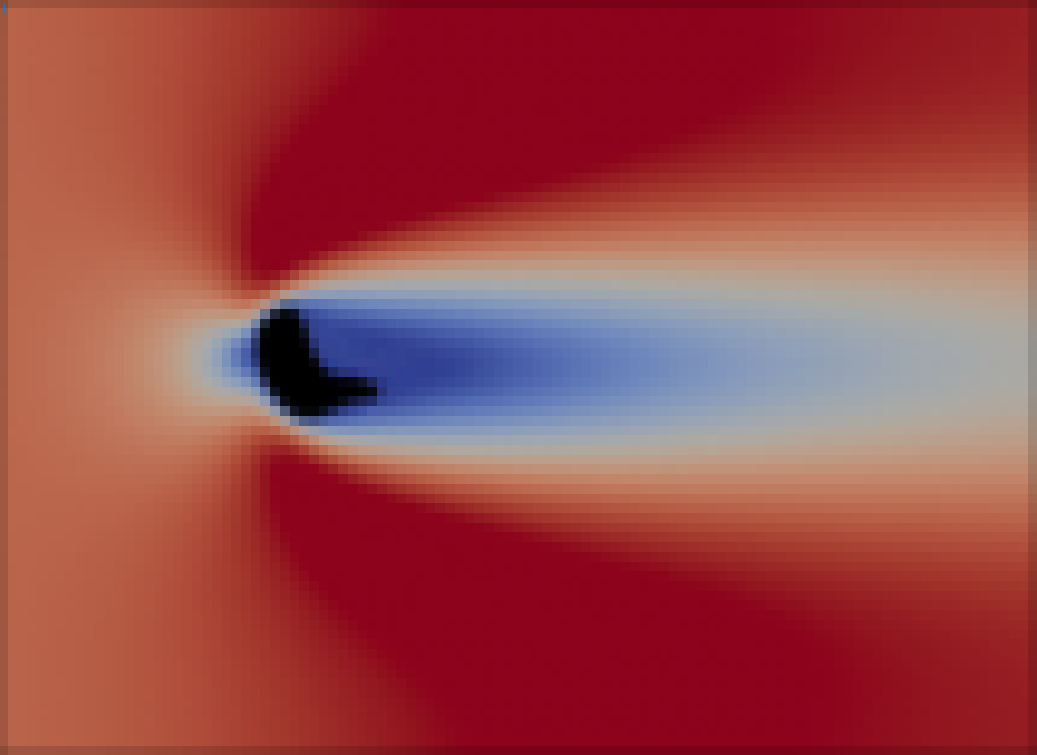}} 
	\caption{Velocity prediction}
	\label{fig:velocity_prediction}
\end{subfigure} \quad
\begin{subfigure}{.45\textwidth}
	\centering
	\fbox{\includegraphics[width=\scale\linewidth]{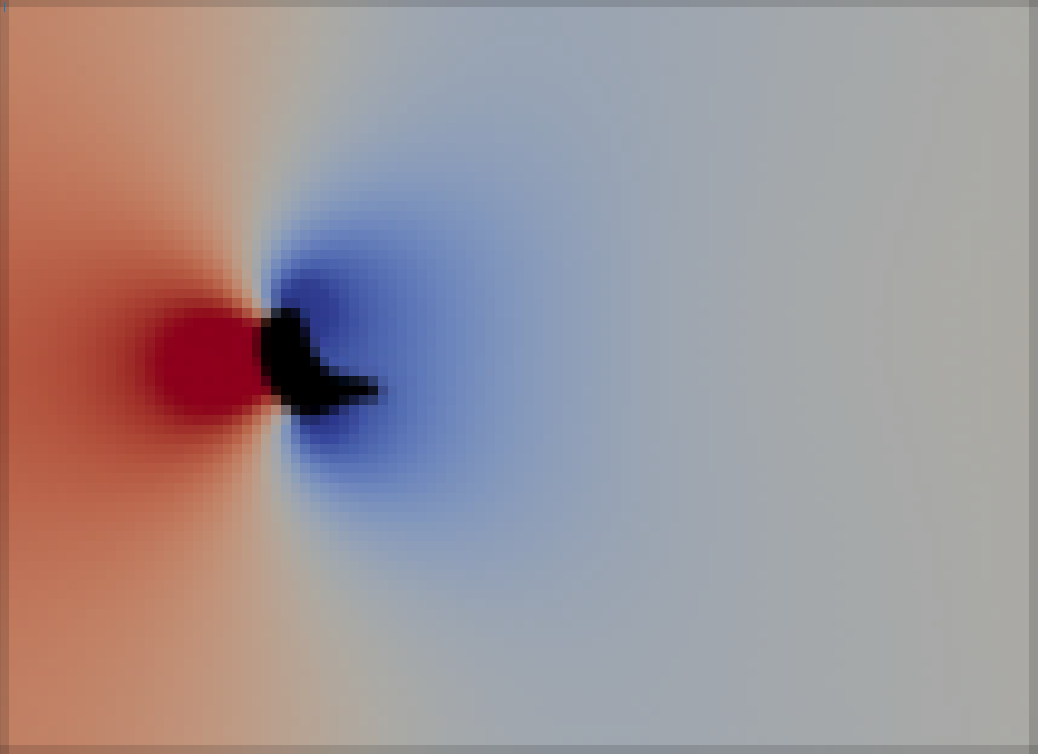}} 
	\caption{Pressure prediction}
	\label{fig:pressure_prediction}
\end{subfigure}

\bigskip

\begin{subfigure}{.45\textwidth}
	\centering
	\fbox{\includegraphics[width=\scale\linewidth]{shape_50_velocity.png}} 
	\caption{Exact velocity}
	\label{fig:velocity_exact}
\end{subfigure} \quad
\begin{subfigure}{.45\textwidth}
	\centering
	\fbox{\includegraphics[width=\scale\linewidth]{shape_50_pressure.png}} 
	\caption{Exact pressure}
	\label{fig:pressure_exact}
\end{subfigure}

\bigskip

\begin{subfigure}{.45\textwidth}
	\centering
	\fbox{\includegraphics[width=\scale\linewidth]{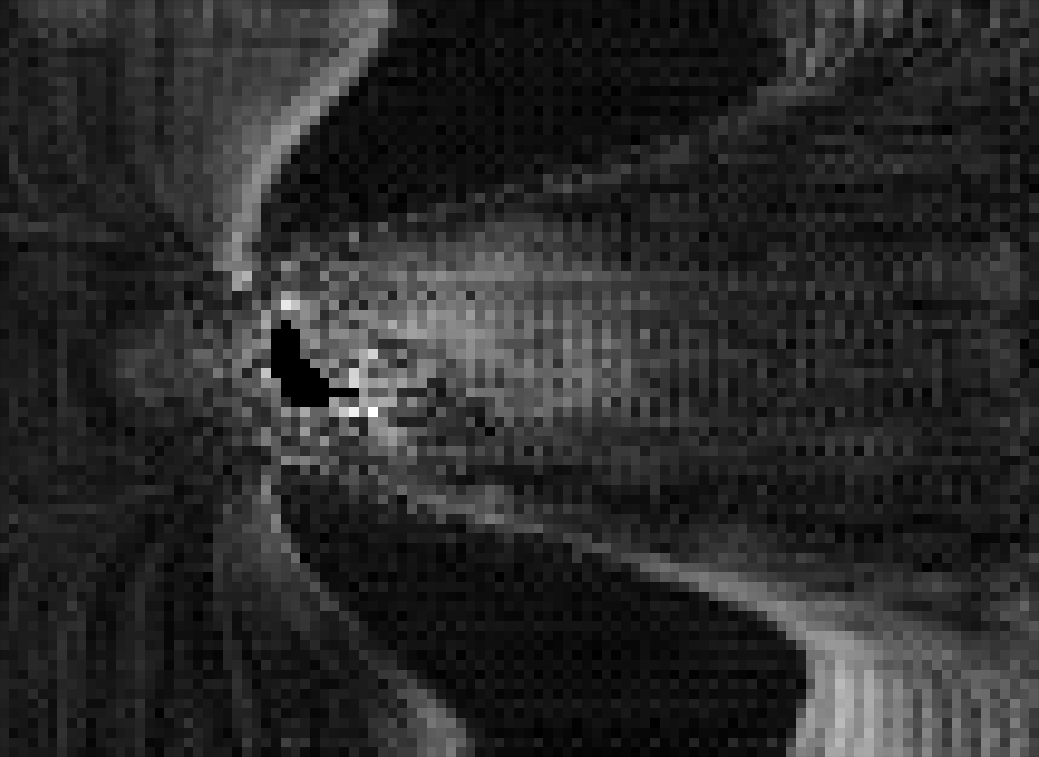}} 
	\caption{Absolute error for velocity prediction}
	\label{fig:velocity_error_abs}
\end{subfigure} \quad
\begin{subfigure}{.45\textwidth}
	\centering
	\fbox{\includegraphics[width=\scale\linewidth]{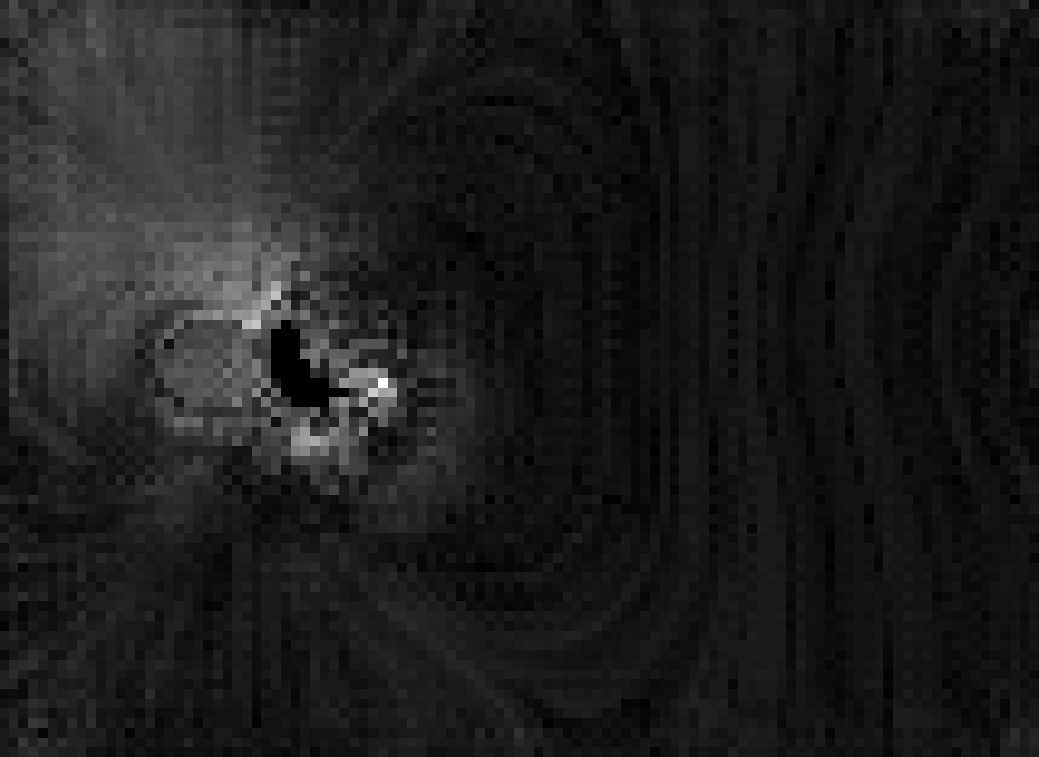}} 
	\caption{Absolute error for pressure prediction}
	\label{fig:pressure_error_abs}
\end{subfigure}

\bigskip

\begin{subfigure}{.45\textwidth}
	\centering
	\fbox{\includegraphics[width=\scale\linewidth]{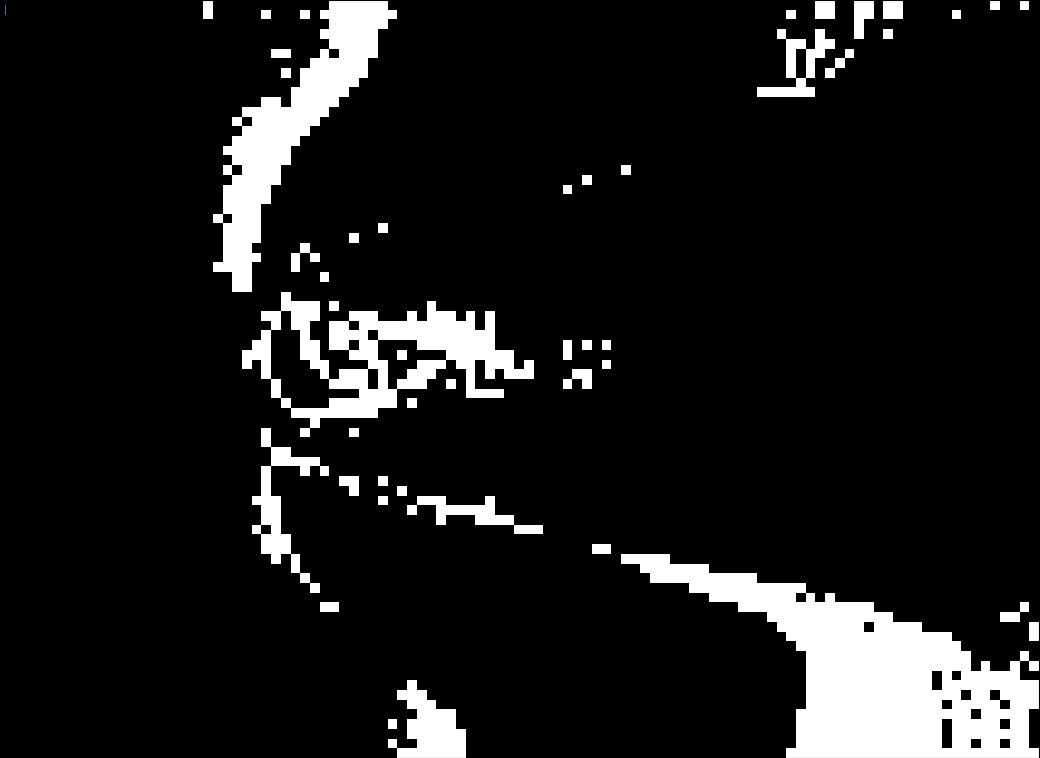}} 
	\caption{Pixels with failed velocity prediction for $\alpha = 0.01$}
	\label{fig:velocity_error_alpha_001}
\end{subfigure} \quad
\begin{subfigure}{.45\textwidth}
	\centering
	\fbox{\includegraphics[width=\scale\linewidth]{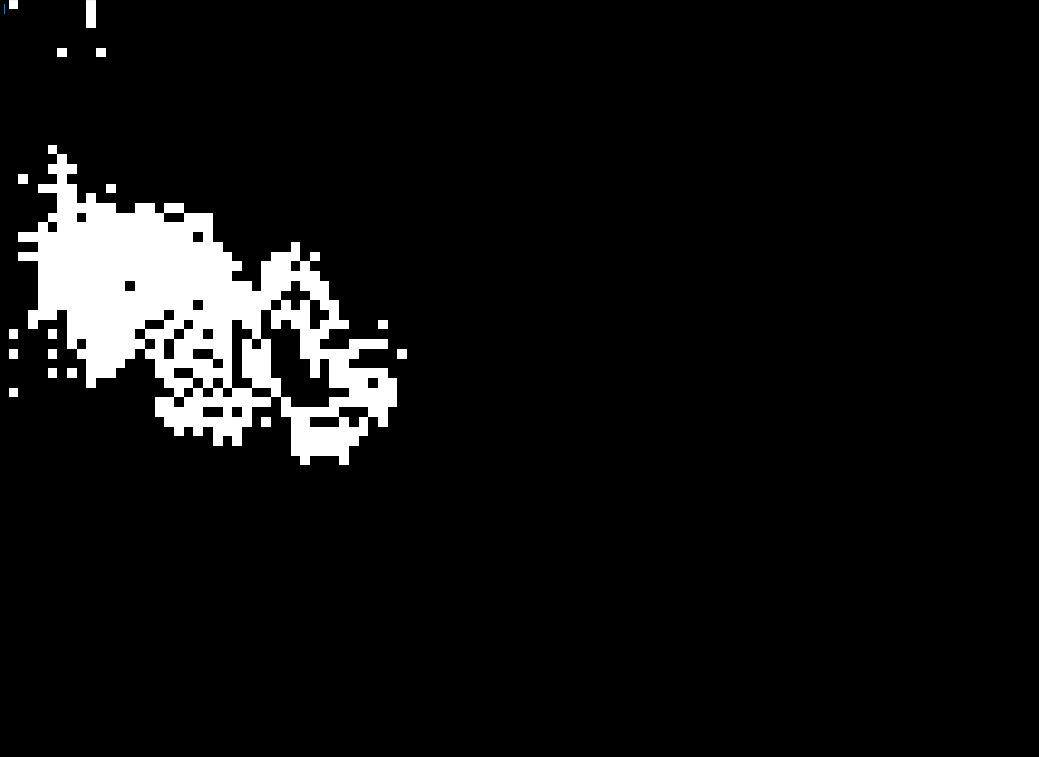}} 
	\caption{Pixels with failed pressure prediction for $\alpha = 0.01$}
	\label{fig:pressure_error_alpha_001}
\end{subfigure}

\bigskip

\begin{subfigure}{.45\textwidth}
	\centering
	\fbox{\includegraphics[width=\scale\linewidth]{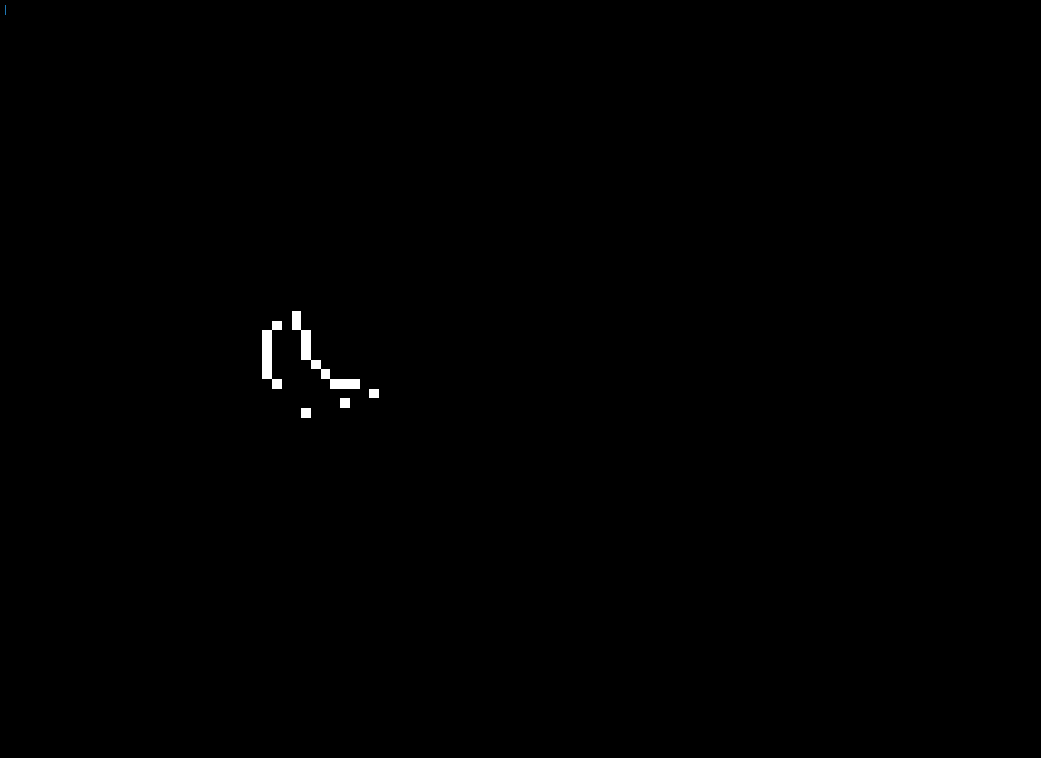}} 
	\caption{Pixels with failed velocity prediction for $\alpha = 0.05$}
	\label{fig:velocity_error_alpha_005}
\end{subfigure} \quad
\begin{subfigure}{.45\textwidth}
	\centering
	\fbox{\includegraphics[width=\scale\linewidth]{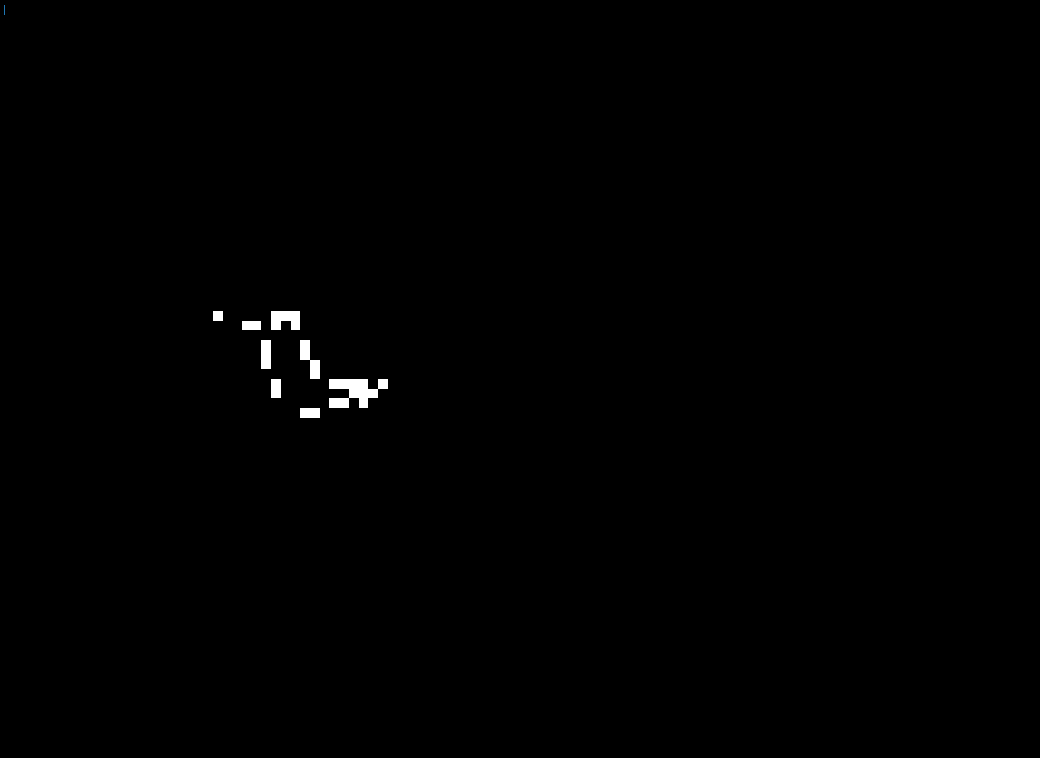}} 
	\caption{Pixels with failed pressure prediction for $\alpha = 0.05$}
	\label{fig:pressure_error_alpha_005}
\end{subfigure}

\caption{\textbf{Velocity (top left) and pressure (top right) predictions on a shape from the test subset, along with their exact solutions and errors to CFD solution}. Fields are scaled to $\left[ 0, 1 \right]$ for velocity and $\left[ -1, 1 \right]$ for pressure, while absolute errors are normalized to their maximal value and scaled to $\left[0, 1 \right]$.}
\label{fig:predictions}
\end{figure}

A prediction example drawn from the test subset is shown in figure \ref{fig:predictions}, along with its corresponding exact solutions and error maps. As can be seen, the predictions are physically sound, and, to the eye, fairly close to their targets. Looking at the pressure predictions, we observe that the network successfully predicted two minimal pressure areas at the top and bottom right of the shape. The maximal errors are located in the direct vicinity of the shape, and in areas of high gradients. By comparing the two error maps $\eps_{0.01}^p$ and $\eps_{0.05}^p$, one sees that most pixels are predicted within a 5\% relative error range, and that the remaining pixels mostly concentrate at the surface of the shape. These error levels could be reduced by reducing the inference region of the network.

\begin{figure}
\centering
\begin{subfigure}{.3\textwidth}
	\centering
	\includegraphics[width=.9\textwidth]{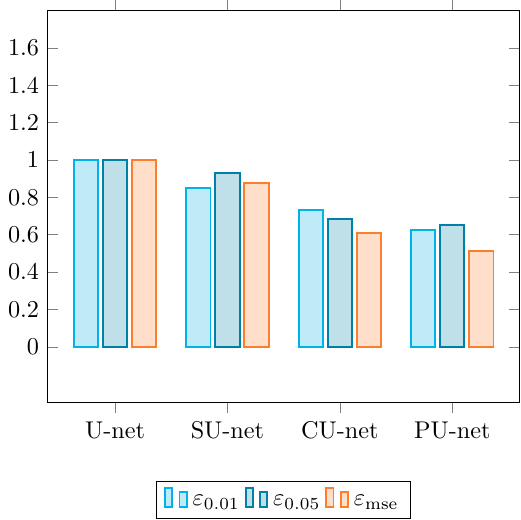}
	\caption{Average error (normalized to U-net) for velocity predictions on test subset}
	\label{fig:avg_error_velocity}
\end{subfigure} \qquad
\begin{subfigure}{.3\textwidth}
	\centering
	\includegraphics[width=.9\textwidth]{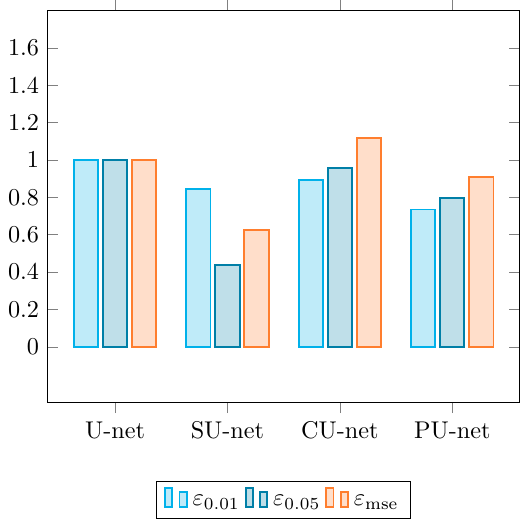}
	\caption{Standard deviation (normalized to U-net) for velocity predictions on test subset}
	\label{fig:avg_deviation_velocity}
\end{subfigure} \qquad
\begin{subfigure}{.3\textwidth}
	\centering
	\includegraphics[width=.9\textwidth]{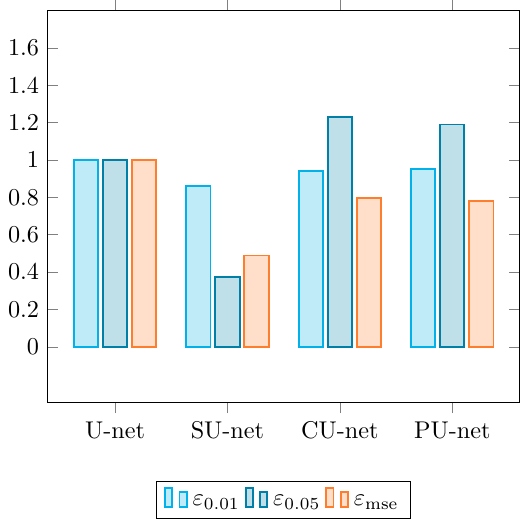}
	\caption{Maximum error (normalized to U-net) for velocity predictions on test subset}
	\label{fig:avg_deviation_velocity}
\end{subfigure}

\bigskip

\begin{subfigure}{.3\textwidth}
	\centering
	\includegraphics[width=.9\textwidth]{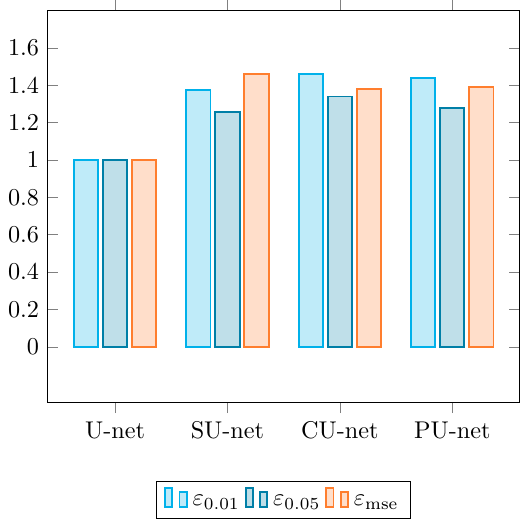}
	\caption{Average error (normalized to U-net) for pressure predictions on test subset}
	\label{fig:avg_error_velocity}
\end{subfigure} \qquad
\begin{subfigure}{.3\textwidth}
	\centering
	\includegraphics[width=.9\textwidth]{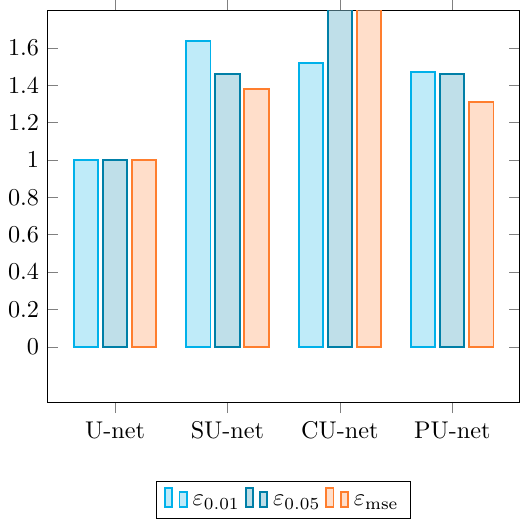}
	\caption{Standard deviation (normalized to U-net) for pressure predictions on test subset}
	\label{fig:avg_deviation_velocity}
\end{subfigure} \qquad
\begin{subfigure}{.3\textwidth}
	\centering
	\includegraphics[width=.9\textwidth]{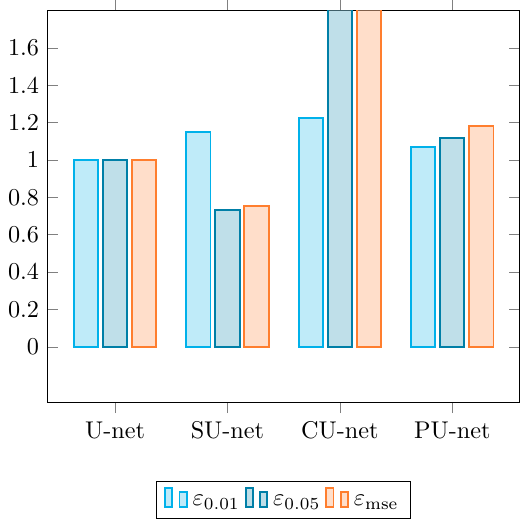}
	\caption{Maximum error (normalized to U-net) for pressure predictions on test subset}
	\label{fig:avg_deviation_velocity}
\end{subfigure}
\caption{\textbf{Average errors, deviations and maximal errors obtained on the test subset using different U-net-based architectures} for velocity (top row) and pressure fields (bottom row).}
\label{fig:prediction_errors}
\end{figure}

As both $\eps_\text{mse}$ and  $\eps_\alpha$ are related to a single image, we compute their average, standard deviation and maximal values to represent these estimators over the entire test subset. The different errors, shown in figure \ref{fig:prediction_errors}, are all normalized to their baseline U-net counterpart for better comparison. On the top row, the error levels for the velocity predictions indicate that, in average, advanced compositions of U-nets provide better predictions in average, up to 45\% of improvement for the PU-net in the MSE sense. Regarding standard deviation and maximum error, SU-nets present a significant advantage, with a drop close to 40\% for standard deviation and 50\% for maximum error using MSE. These results suggest that SU-nets are able to catch outliers with more efficiency than CU- and PU-nets. 

Although one could expect to observe similar results for pressure predictions, the performances of SU-, CU- and PU-nets are clearly below that of the baseline U-net for average error, standard deviation and maximum error. The fact that pressure predictions cannot be improved using more complex networks can arise from an insufficient pre-processing of the target data, causing the learning problem to be ill-conditioned, as suggested in \cite{Thuerey2018}. Still, the improvements obtained on the velocity predictions when using more complex networks may not be worthwhile in regard of the overhead of computational effort they require. Indeed, the baseline U-net shown in figure \ref{fig:Unet} contains 7.7 millions parameters, and requires 3 hours of training on a regular computer (not using GPU cards), while the SU-, CU- and PU-nets contain between 15 and 18 million parameters, leading to training times of 6 to 7 hours. For that reason, we suggest that, although improved U-net-based architectures provide better results in segmentation tasks, such superiority may not reflect in regression tasks such as field map predictions.

\begin{figure}
\centering

\setlength{\fboxsep}{0pt}%
\setlength{\fboxrule}{1pt}%

\def\scale{.175}

\begin{subfigure}{\scale\textwidth}
	\centering
	\fbox{\includegraphics[width=\linewidth]{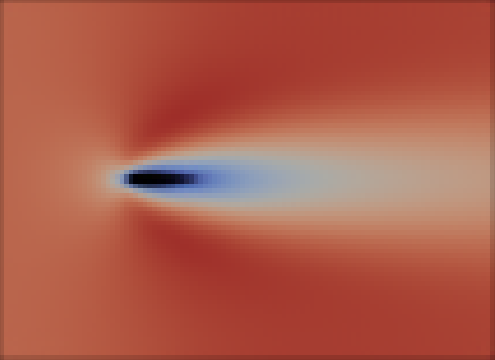}} 
\end{subfigure} \quad
\begin{subfigure}{\scale\textwidth}
	\centering
	\fbox{\includegraphics[width=\linewidth]{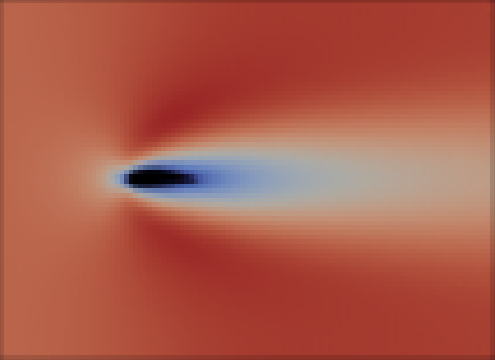}} 
\end{subfigure} \quad
\begin{subfigure}{\scale\textwidth}
	\centering
	\fbox{\includegraphics[width=\linewidth]{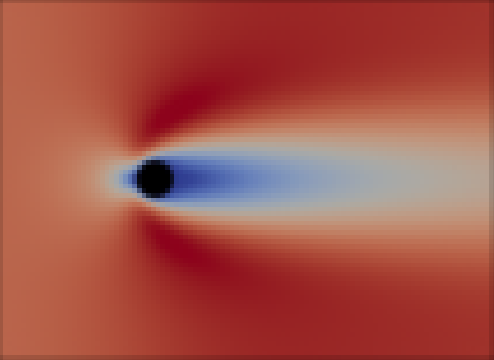}} 
\end{subfigure} \quad
\begin{subfigure}{\scale\textwidth}
	\centering
	\fbox{\includegraphics[width=\linewidth]{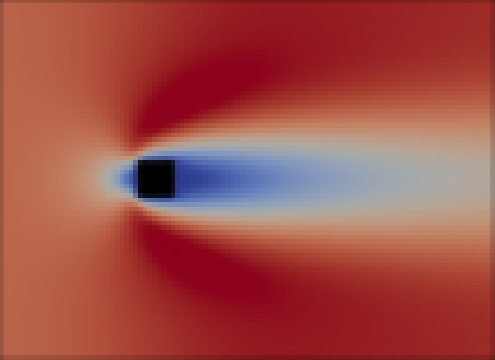}} 
\end{subfigure} \quad
\begin{subfigure}{\scale\textwidth}
	\centering
	\fbox{\includegraphics[width=\linewidth]{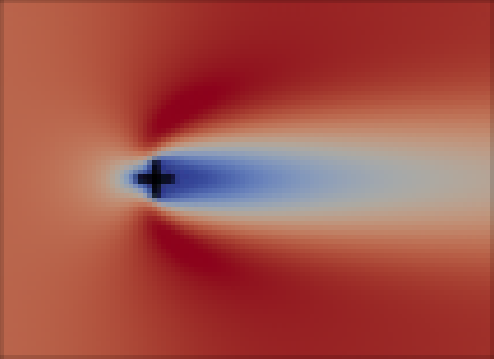}} 
\end{subfigure} 

\medskip

\begin{subfigure}{\scale\textwidth}
	\centering
	\fbox{\includegraphics[width=\linewidth]{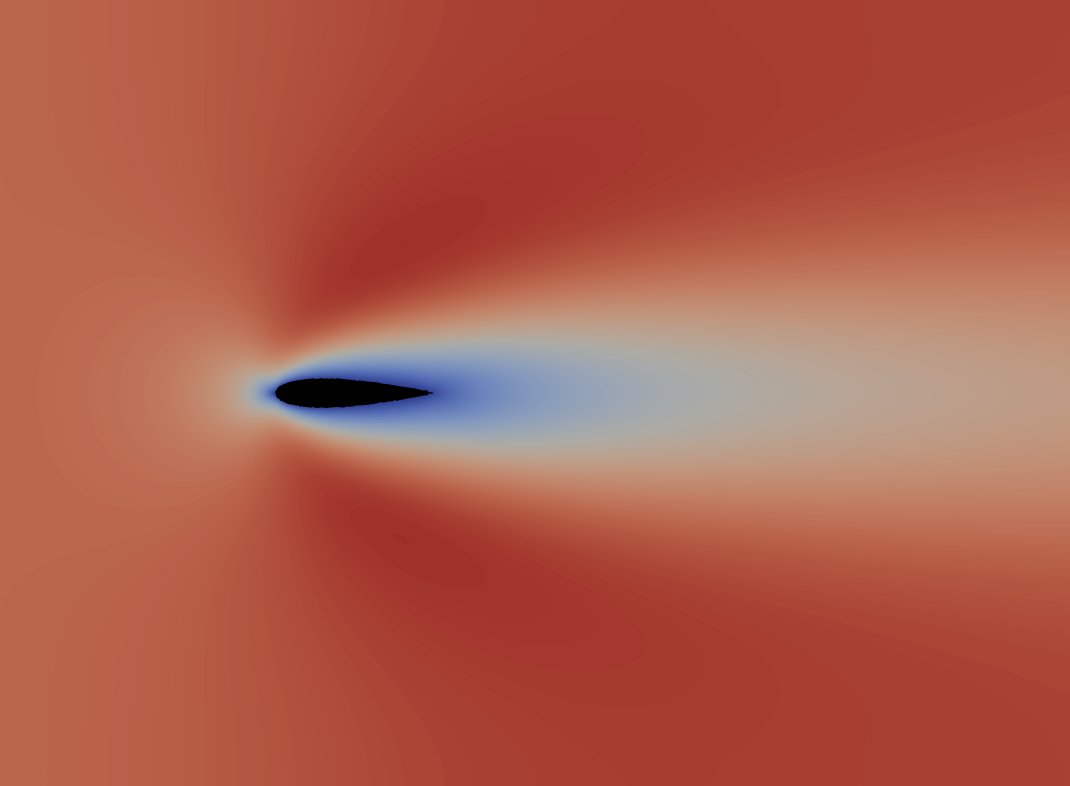}} 
\end{subfigure} \quad
\begin{subfigure}{\scale\textwidth}
	\centering
	\fbox{\includegraphics[width=\linewidth]{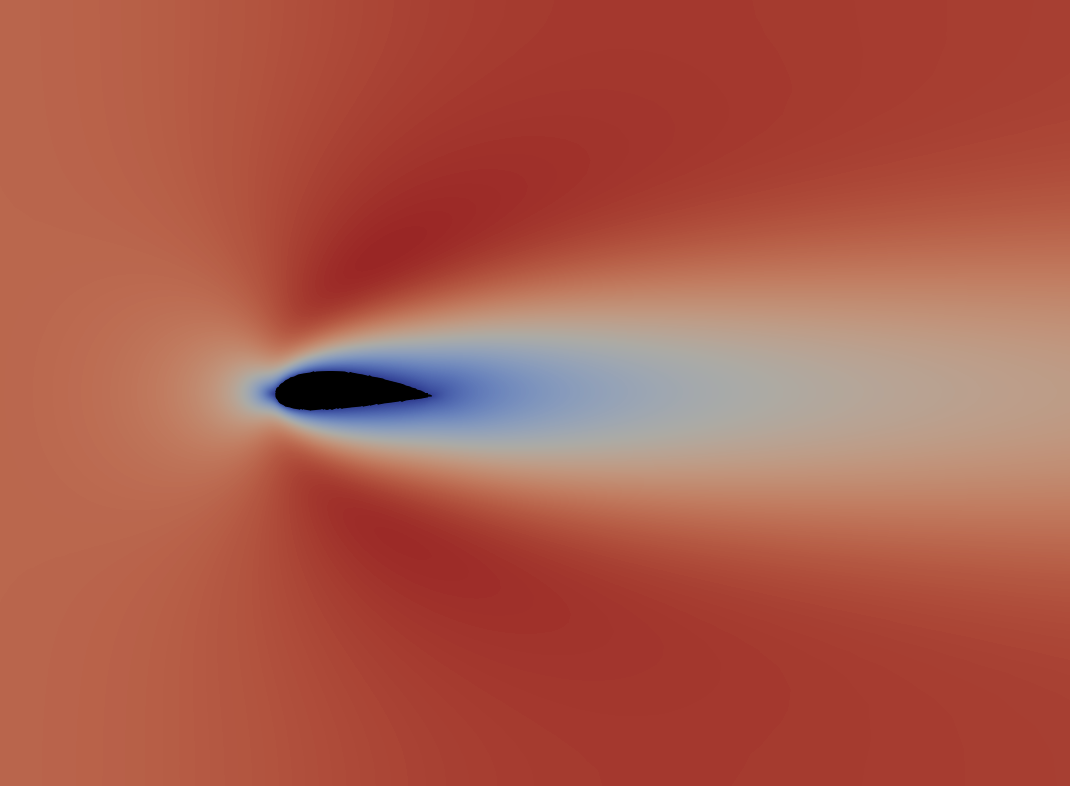}} 
\end{subfigure} \quad
\begin{subfigure}{\scale\textwidth}
	\centering
	\fbox{\includegraphics[width=\linewidth]{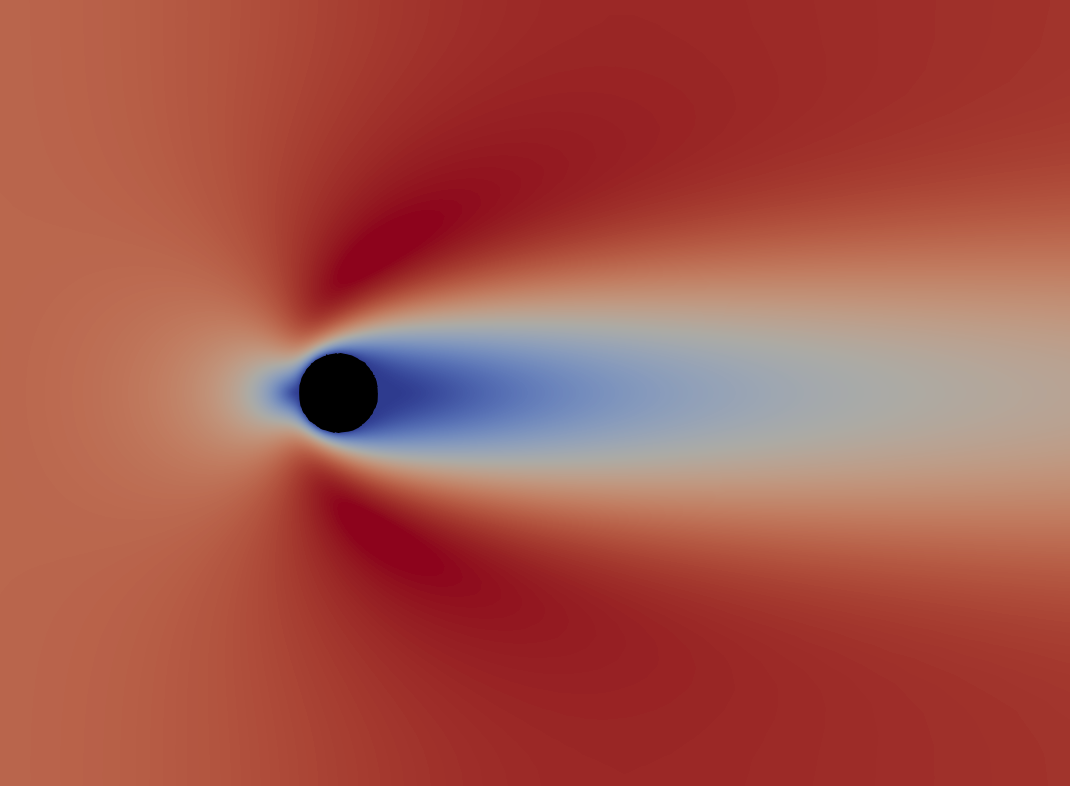}} 
\end{subfigure} \quad
\begin{subfigure}{\scale\textwidth}
	\centering
	\fbox{\includegraphics[width=\linewidth]{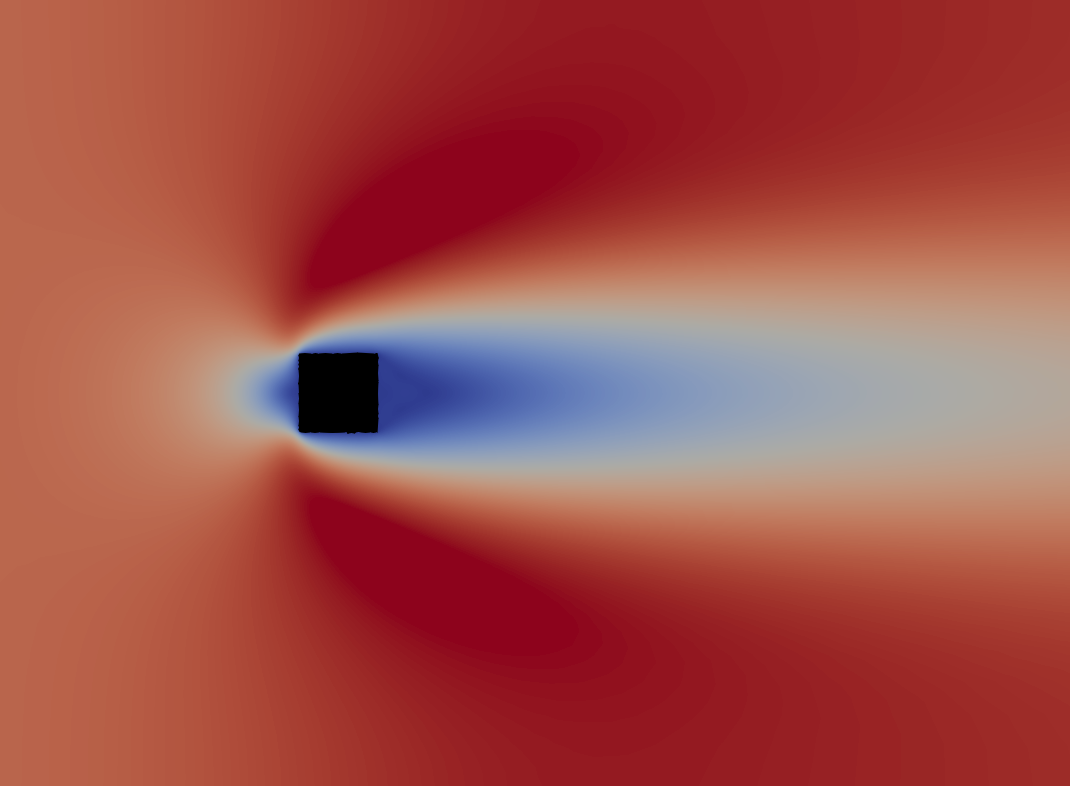}} 
\end{subfigure} \quad
\begin{subfigure}{\scale\textwidth}
	\centering
	\fbox{\includegraphics[width=\linewidth]{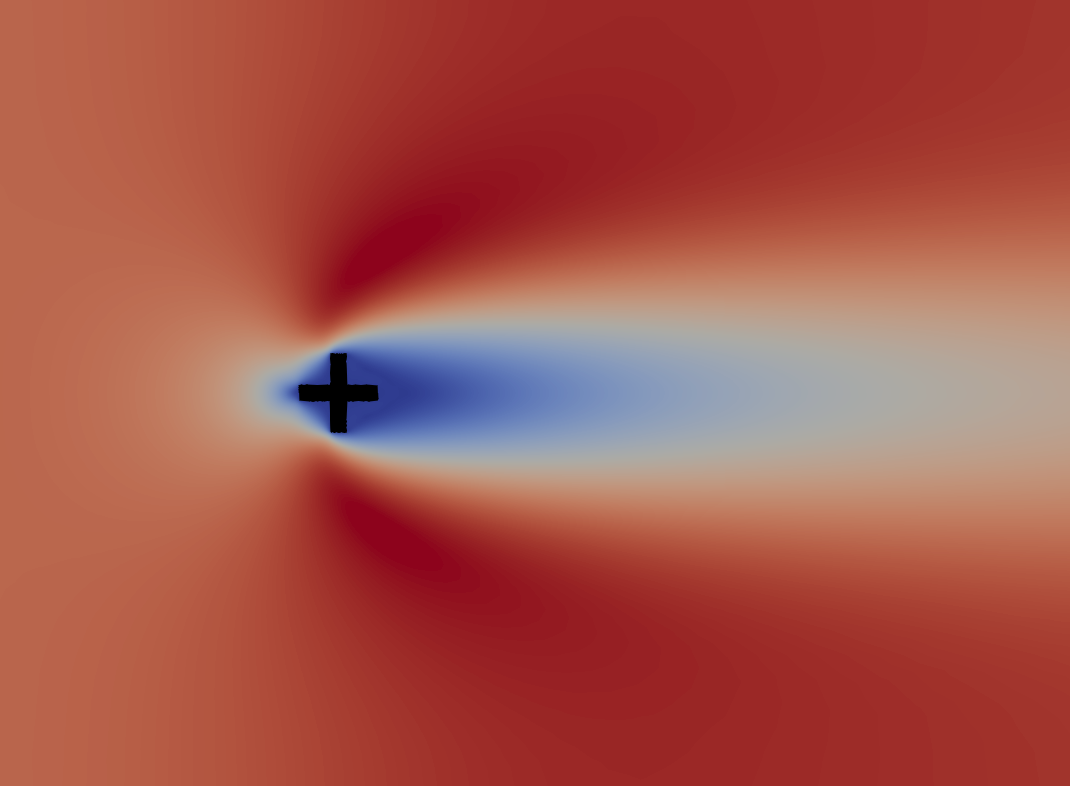}} 
\end{subfigure}

\medskip

\begin{subfigure}{\scale\textwidth}
	\centering
	\fbox{\includegraphics[width=\linewidth]{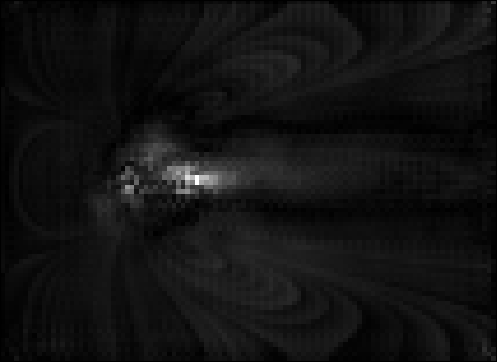}} 
\end{subfigure} \quad
\begin{subfigure}{\scale\textwidth}
	\centering
	\fbox{\includegraphics[width=\linewidth]{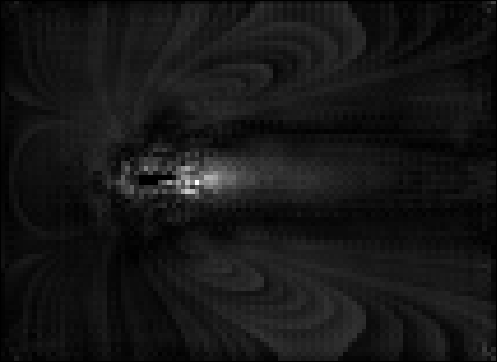}} 
\end{subfigure} \quad
\begin{subfigure}{\scale\textwidth}
	\centering
	\fbox{\includegraphics[width=\linewidth]{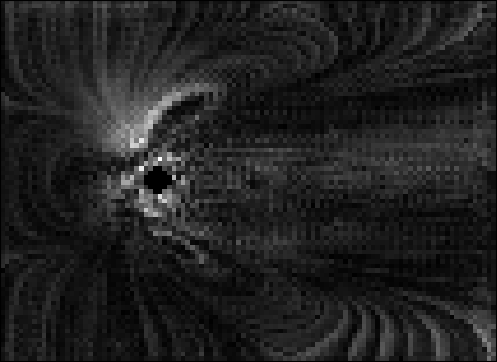}} 
\end{subfigure} \quad
\begin{subfigure}{\scale\textwidth}
	\centering
	\fbox{\includegraphics[width=\linewidth]{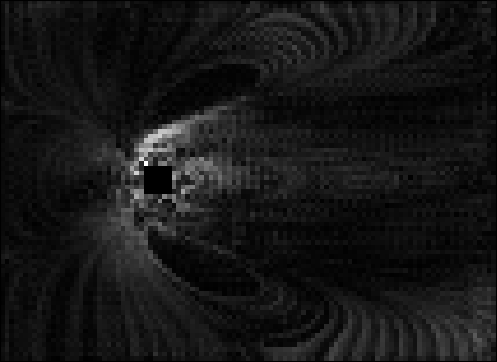}} 
\end{subfigure} \quad
\begin{subfigure}{\scale\textwidth}
	\centering
	\fbox{\includegraphics[width=\linewidth]{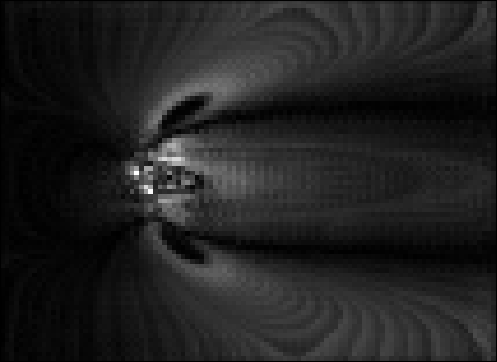}} 
\end{subfigure}

\medskip

\begin{subfigure}{\scale\textwidth}
	\centering
	\fbox{\includegraphics[width=\linewidth]{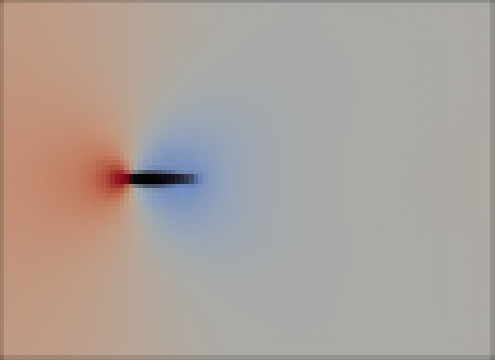}} 
\end{subfigure} \quad
\begin{subfigure}{\scale\textwidth}
	\centering
	\fbox{\includegraphics[width=\linewidth]{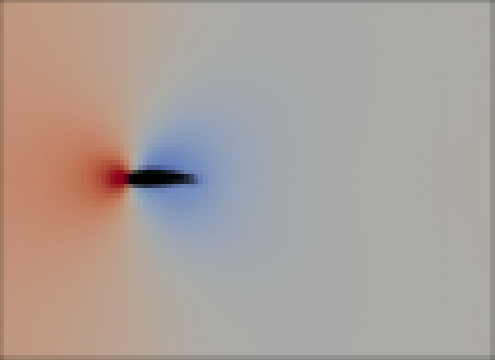}} 
\end{subfigure} \quad
\begin{subfigure}{\scale\textwidth}
	\centering
	\fbox{\includegraphics[width=\linewidth]{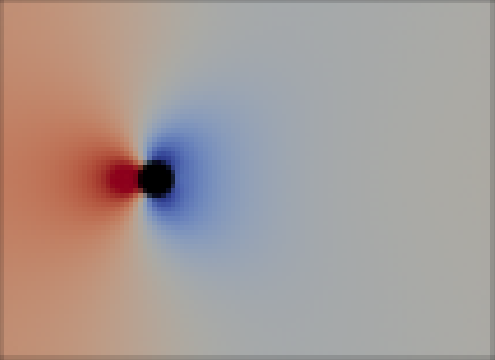}} 
\end{subfigure} \quad
\begin{subfigure}{\scale\textwidth}
	\centering
	\fbox{\includegraphics[width=\linewidth]{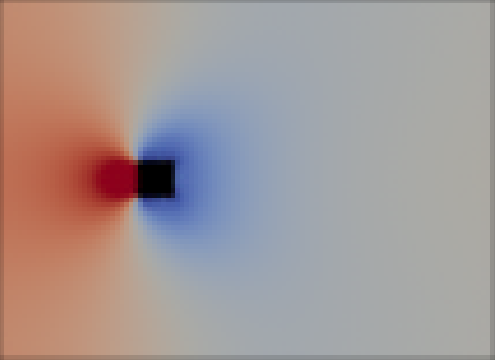}} 
\end{subfigure} \quad
\begin{subfigure}{\scale\textwidth}
	\centering
	\fbox{\includegraphics[width=\linewidth]{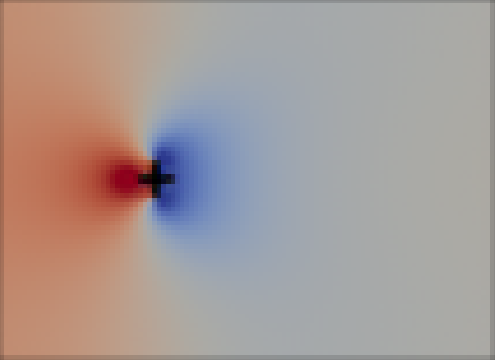}} 
\end{subfigure}

\medskip

\begin{subfigure}{\scale\textwidth}
	\centering
	\fbox{\includegraphics[width=\linewidth]{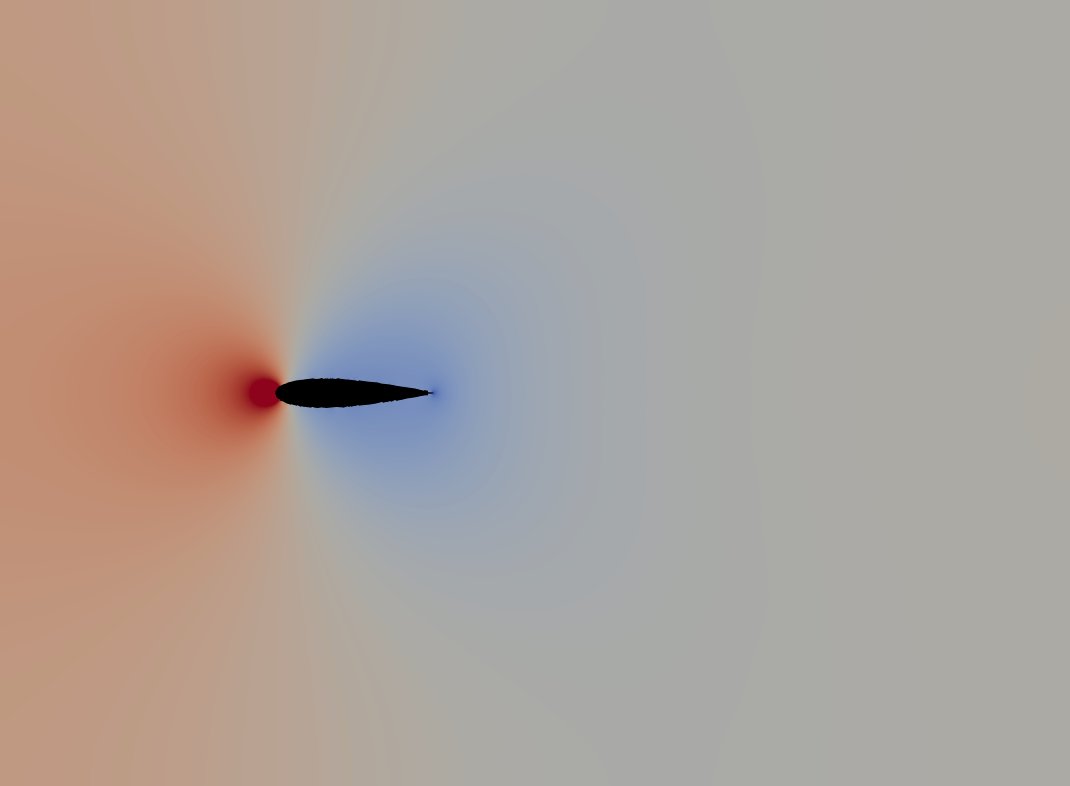}} 
\end{subfigure} \quad
\begin{subfigure}{\scale\textwidth}
	\centering
	\fbox{\includegraphics[width=\linewidth]{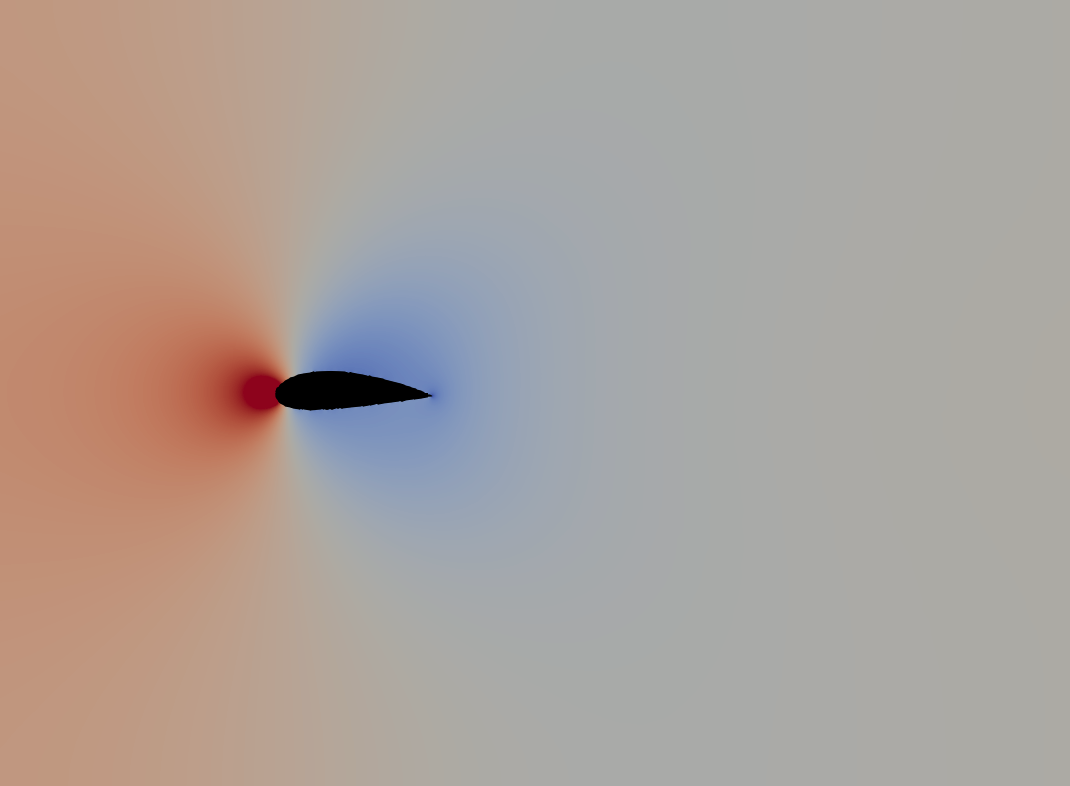}} 
\end{subfigure} \quad
\begin{subfigure}{\scale\textwidth}
	\centering
	\fbox{\includegraphics[width=\linewidth]{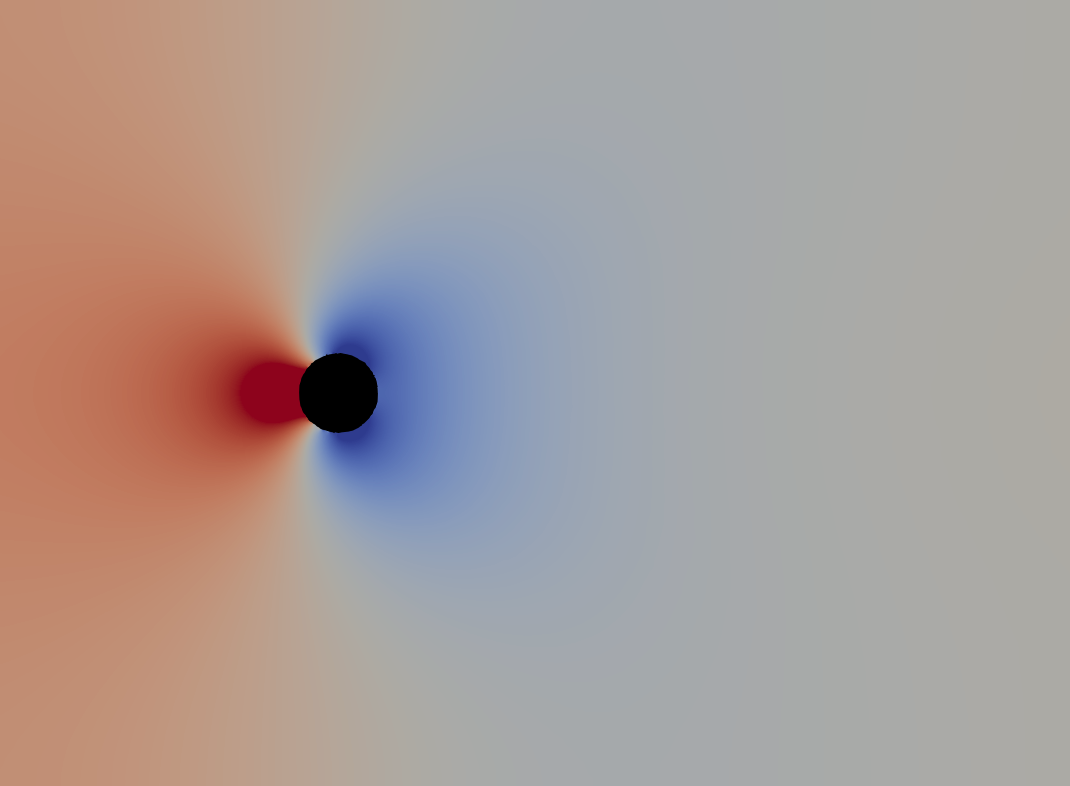}} 
\end{subfigure} \quad
\begin{subfigure}{\scale\textwidth}
	\centering
	\fbox{\includegraphics[width=\linewidth]{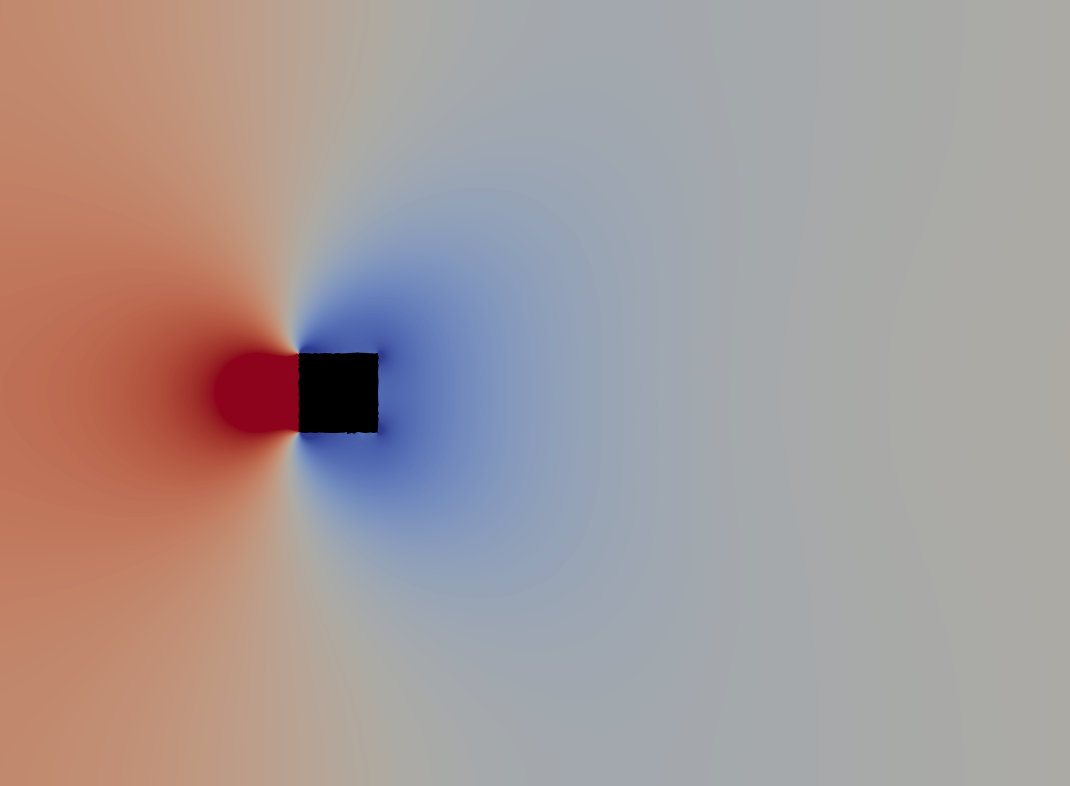}} 
\end{subfigure} \quad
\begin{subfigure}{\scale\textwidth}
	\centering
	\fbox{\includegraphics[width=\linewidth]{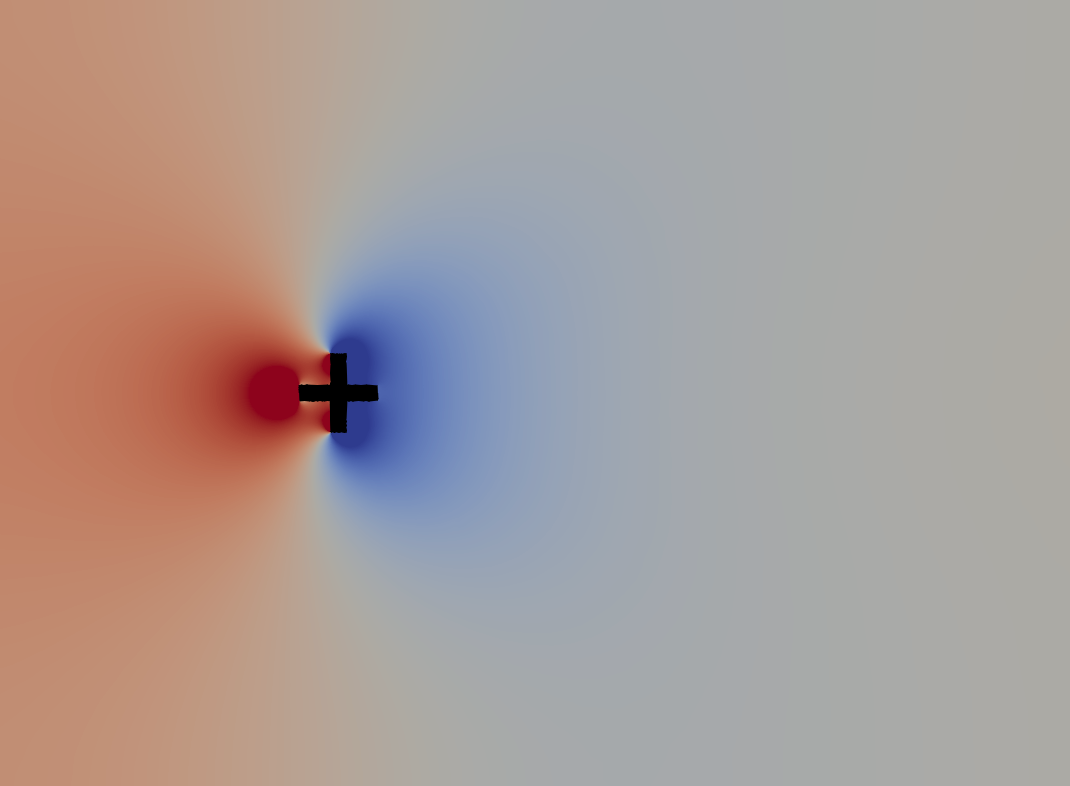}} 
\end{subfigure}

\medskip

\begin{subfigure}{\scale\textwidth}
	\centering
	\fbox{\includegraphics[width=\linewidth]{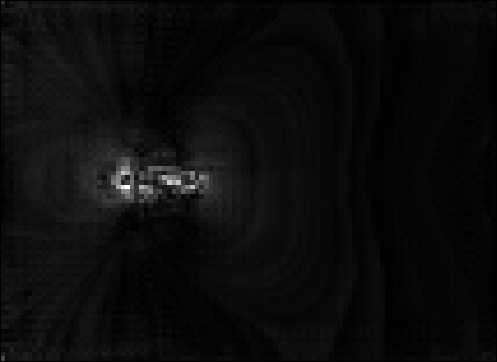}} 
\end{subfigure} \quad
\begin{subfigure}{\scale\textwidth}
	\centering
	\fbox{\includegraphics[width=\linewidth]{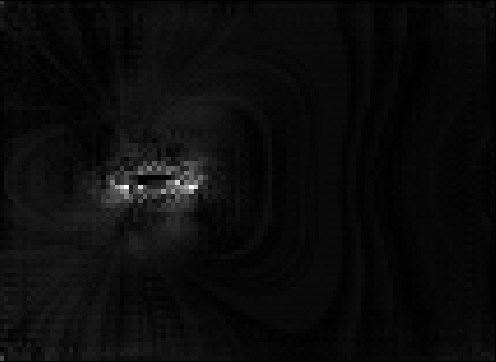}} 
\end{subfigure} \quad
\begin{subfigure}{\scale\textwidth}
	\centering
	\fbox{\includegraphics[width=\linewidth]{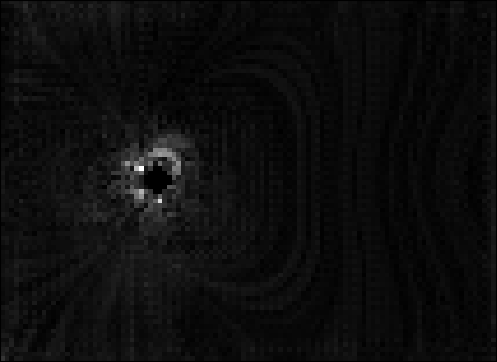}} 
\end{subfigure} \quad
\begin{subfigure}{\scale\textwidth}
	\centering
	\fbox{\includegraphics[width=\linewidth]{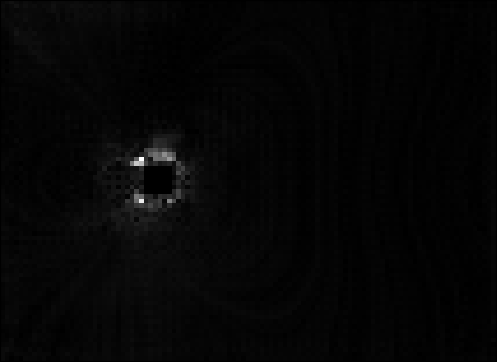}} 
\end{subfigure} \quad
\begin{subfigure}{\scale\textwidth}
	\centering
	\fbox{\includegraphics[width=\linewidth]{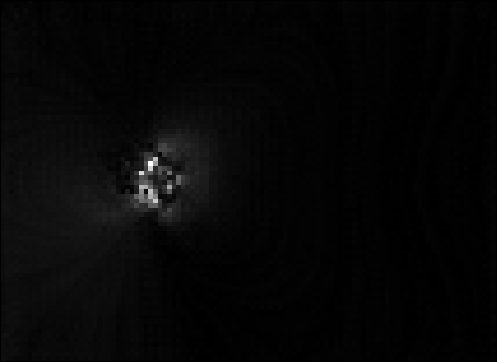}} 
\end{subfigure}

\caption{\textbf{Velocity (first row) and pressure (fourth row) predictions compared to their exact counterparts (second and fifth rows), along with their associated absolute error maps (third and sixth rows) for different real-life shapes.}. From left to right: NACA 0018, NACA 4424, cylinder, square and cross.}
\label{fig:predictions_unseen}
\end{figure}

We now focus on the prediction of field maps in realistic, unseen configurations such as geometric shapes or NACA airfoils, using the baseline U-net architecture of figure \ref{fig:Unet}. As shown in figure \ref{fig:predictions_unseen}, predictions provided by the trained U-net are still physically sound, and field details such as multiple high and low pressure areas in the vicinity of the obstacles are well reproduced, to the limit of the resolution of the network output. MSE error levels for these predictions are in the same order of magnitude as those obtained on the test subset, \ie between \num{1e-5} and \num{5e-5} for both pressure and velocity. Further testing would require to reduce the predicted domain around the shapes to obtain better details in its vicinity, or to output predictions with higher resolution, although it would require more computational power.

\section{Conclusion}

In the current contribution, we presented the prediction of velocity and pressure field maps around arbitrary 2D shapes in laminar flows exploiting U-net-based architectures. First, details were given about the considered U-net architectures, and how they recently started to be exploited for regression tasks. Insights were then given on the dataset construction and the networks training, followed by a discussion on the metrics used to measure the quality of the predictions. A prediction example was then analyzed on a shape drawn from the test subset, that showed good agreement with the exact solution for both velocity and pressure. Then, the predictions were assessed statistically on the entire test subset, showing that, although they perform better on segmentation tasks, advanced architectures such as stacked, coupled and parallel U-nets do not provide a clear advantage on regression tasks such as field map predictions. Finally, we showed that such predictions could be made on unseen shapes with reasonable accuracy, and that the predictions were physically sound, thanks to the rich geometrical features present in the dataset. 

Although this application is focused on a fairly basic case, these results give hope for the possibility of tackling more complex problems, such as turbulent viscosity maps at higher Reynolds numbers, or 3D fields, for example. Additionally, such tools can be exploited in related domains as surrogate models, for optimization purposes.

\bibliographystyle{elsarticle-num} 
\biboptions{sort&compress}
\bibliography{nn_flow_2d}

\end{document}